\documentclass{article}

\usepackage[a4paper,left=2cm,right=2cm,top=2.5cm,bottom=2.5cm]{geometry}
\usepackage{graphicx}
\usepackage{amsmath}
\usepackage{amssymb}
\usepackage{verbatim}
\usepackage{multirow}
\usepackage{tikz}
\usetikzlibrary{quantikz2}
\usepackage{float}
\usepackage{subcaption}
\usepackage{array}
\usepackage{pgfplots}
\pgfplotsset{compat=1.18}
\usepackage{amsthm}
\usepackage{url}
\usepackage{hyperref}
\usepackage{xcolor}
\hypersetup{colorlinks=false,allbordercolors=white}
\usepackage{algorithmic}
\usepackage{algorithm}
\usepackage{authblk}
\usepackage{cancel}
\usepackage{booktabs}

\title{Algebraic Reduction to Improve an Optimally Bounded Quantum State Preparation Algorithm}

\author{Giacomo Belli and Michele Amoretti}

\affil{Quantum Software Laboratory, Department of Engineering and Architecture\\University of Parma, 43124 Parma, Italy}

\date{}

\begin{document}

\maketitle

\begin{abstract}
The preparation of $n$-qubit quantum states is a cross-cutting subroutine for many quantum algorithms, and the effort to reduce its circuit complexity is a significant challenge. In the literature, the quantum state preparation algorithm by Sun et al. is known to be optimally bounded, defining the asymptotically optimal width-depth trade-off bounds with and without ancillary qubits. In this work, a simpler algebraic decomposition is proposed to separate the preparation of the real part of the desired state from the complex one, resulting in a reduction in terms of circuit depth, total gates, and CNOT count when $m$ ancillary qubits are available.
The reduction in complexity is due to the use of a single operator $\Lambda$ for each uniformly controlled gate, instead of the three in the original decomposition. Using the PennyLane library, this new algorithm for state preparation has been implemented and tested in a simulated environment for both dense and sparse quantum states, including those that are random and of physical interest. Furthermore, its performance has been compared with that of M\"ott\"onen et al.'s algorithm, which is a de facto standard for preparing quantum states in cases where no ancillary qubits are used, highlighting interesting lines of development.

\textbf{keywords} - \textit{Quantum state preparation, Ancillary qubits, Fourier space, Diagonal operators}
\end{abstract}


\section{Introduction}

Quantum state preparation (QSP) is a fundamental subroutine in many quantum algorithms~\cite{Iaconis2024}. From quantum linear system solvers~\cite{harrow2009quantum,morales2024quantum,bravo2023variational} to Hamiltonian simulation~\cite{berry2018improved, low2019hamiltonian, babbush2018encoding}, from quantum chemistry~\cite{mcardle2020quantum,fomichev2024initial,wang2022state,berry2025rapid}  to quantum machine learning~\cite{rath2024quantum,caro2021encoding}, the constant search for more efficient algorithms is still an open challenge~\cite{gui2024spacetime,malz2024preparation,mao2024toward,sood2023towards,lee2023evaluating,cao2019quantum,ward2009preparation}. The goal of a QSP algorithm is to construct a unitary operator $U\in U(2^n)$ such that $|\psi\rangle=U|0\rangle^{\otimes n}$, where $|0\rangle^{\otimes n}$ is the conventional $n$-qubit initial state and $|\psi\rangle=\sum_{i=0}^{2^n-1}c_i|i\rangle$ is the desired quantum state in the computational basis, normalized with respect to the $l_2$-norm. The preparation can follow an exact or approximate paradigm; in the approximate version, one tries to prepare a state $|\phi\rangle$ that is $\epsilon$-close to $|\psi\rangle$ with respect to some metric. Furthermore, the QSP algorithm can benefit from the use of ancillary registers in addition to the input registers to decrease some computational costs, such as depth, which corresponds to execution time and gate counts. Indeed, the \emph{ancillae-based} strategy aligns with the industrial effort to scale quantum chips to ever-larger qubit counts, having to deal with intrinsically limited coherence times~\cite{ibm_scaling,ibm_roadmap}. In this industrial context, designing algorithms on larger logical registers via ancillary qubits appears to be a promising direction~\cite{bravyi2024high,khattar2025rise,chatterjee2025quantum}. Clearly, one could try to minimize each cost measure individually, for example, by focusing only on depth. However, of particular interest is the study of the optimal space-time trade-off for quantum circuits, also with respect to hardware connectivity~\cite{yuan2023optimal,yuan2023does}.

The study of QSP began in its exact version and without ancillary qubits using the method proposed by Nielsen and Chuang~\cite{nielsen2000quantum}, characterized by a depth upper bound of $\mathcal{O}(2^n)$. In the same period, Grover and Rudolph~\cite{grover2002creating} proposed a \emph{hierarchical} QSP based on layers of conditional rotations for the special case of superpositions of integrable distributions, with a depth upper bound of $\mathcal{O}(n2^n)$. Their algorithm contained what we might define today as the \emph{embryonic form} of Uniformly Controlled Rotations (UCR)~\cite{mottonen12006decompositions}, even though the UCR class had not yet been formalized. In 2005, M\"ott\"onen et al.~\cite{mottonen2005transformation} proposed a 3-stage QSP that properly formalizes the UCR class, with a depth of $\mathcal{O}(2^n)$, $2^{n+2}-4n-4$ CNOT, and $2^{n+2}-5$ rotations. Each UCR layer can be implemented via Gray Code~\cite{vartiainen2004efficient}, and its parameters can be determined via Binary Search Tree (BST) diagrams, making QSP modular~\cite{mottonen2004quantum}. Immediately after, Bergholm et al.~\cite{bergholm2005quantum}, still within the framework of a general QSP, introduced and systematically analyzed the Uniformly Controlled Gate (UCG or multiplexor) class\footnote{IBM Quantum, Qiskit Documentation: \emph{UCGate class}, https://docs.quantum.ibm.com/api/qiskit/qiskit.circuit.library.UCGate.} as a generalization of the UCR class. Their algorithm had an upper bound of $2^{n+1}-2n-2$ for the number of CNOT gates, with a depth also of order $\mathcal{O}(2^n)$.

Following the line of research on QSP, where the UCG is the key building block and no ancillary qubits are used, from 2006 to 2016, some important results structured a bridge (not only theoretical) between Gate Synthesis and QSP, where efficient general decomposition techniques (QR/CS) are combined with the modularity of UCG/UCR~\cite{mottonen12006decompositions}. While Shende et al.~\cite{shende2006synthesis} proposed a framework for synthesis pipelines, with tighter asymptotic bounds than the literature at the time and the handling of the nearest-neighbor topology, Plesch and Brukner~\cite{plesch2011quantum} focused on minimizing the CNOT count for a general QSP with arbitrary decomposition, achieving $\frac{23}{24}2^n-2^{\frac{n}{2}+1}+\frac{5}{3}$ for even $n$ and $\frac{115}{96}2^n$ for odd $n$. Finally, Iten et al.~\cite{iten2016quantum} provided the framework for the systematic synthesis of isometries, within which QSP represents a special case. Their work analyzed three decompositions close to the theoretical lower bound for the number of CNOT, defining what is still an industrial benchmark\footnote{In IBM's Qiskit library, the implementation of the algorithms for state and isometry synthesis has been unified under the approach proposed by Iten, Colbeck, and Christandl~\cite{iten2016quantum}. In particular, the classes \emph{StatePreparation}, \emph{Isometry}, and \emph{Initialize} (all within the qiskit.circuit.library module) invoke routines derived from this work, which thus became the reference framework for state and isometry construction in industrial quantum compilers. See IBM Quantum Documentation: https://docs.quantum.ibm.com/api/qiskit/qiskit.circuit.library.}. These milestones in the development of QSP clarify the link with unitary synthesis and mark the transition from the preparation of specific quantum states to the design of unitary operators, where QSP can be seen as a sub-routine for preparing columns.

The use of ancillae in the history of QSP began with Zhang et al.~\cite{zhang2021low}, whose algorithm can generate the target state in $\mathcal{O}(n^2)$ depth with $\mathcal{O}(2^{2n})$ ancillary qubits, but only with a limited success probability. In the same year, Rosenthal~\cite{rosenthal2021query} tackled the topic of space-time trade-off by obtaining an algorithm with depth $\mathcal{O}(n)$ using $\mathcal{O}(n2^n)$ ancillary qubits. The \emph{asymptotically} optimal space-time trade-off bounds were found by Sun et al.~\cite{sun2023asymptotically} in their algorithm, hereinafter referred to as SUN, with depth $\mathcal{O}(\frac{2^n}{m+n}+n\log n)$ and width $\mathcal{O}(2^n)$, where $m$ is the number of ancillary qubits. The QSP circuit by Zhang et al.~\cite{zhang2022quantum} of depth $\mathcal{O}(n)$ and using $\Theta(2^n)$ ancillary qubits is a special case of~\cite{sun2023asymptotically}. This line of research culminated in the work of Yuan et al.~\cite{yuan2023optimal}, which completely solves the circuit complexity with or without ancillary qubits.

This paper provides the following contributions: 
\begin{enumerate}
\item a novel QSP algorithm that optimizes SUN in terms of depth and gate counts over the entire first range of~\cite[Theorem 1]{sun2023asymptotically}; 
\item a simpler algebraic decomposition that splits the preparation of the real part of the desired state from the complex one and allows each UCG to be implemented with a single $\Lambda$-type constructor (instead of the 3 of the original version~\cite{sun2023asymptotically}); 
\item a tighter asymptotic optimal bound for the space-time trade-off in the first range of~\cite[Theorem 1]{sun2023asymptotically}; 
\item the implementation of the new algorithm via the PennyLane library, available on the GitHub repository~\cite{QSP-Sun}; 
\item a complete verification of the algorithm's performance through an extensive battery of tests (sparse and dense states, with complex or real coefficients, and states of interest for research) executed in a simulated environment on the HPC system of the University of Parma; 
\item an explicit comparison is made in terms of depth, total gates, CNOT, and gate type with Mottonen's QSP algorithm (the standard without ancillae) and with the original version by Sun et al. (the standard for optimal complexity).
\end{enumerate}

The remainder of the paper is organized as follows. In Section~\ref{sec:qsp_sun}, the QSP algorithm by Sun et al.~\cite{sun2023asymptotically} is presented with its essential features, thus providing the theoretical framework and the main equations to understand its optimization. In Section~\ref{sec:new_decomposition}, the new decomposition that splits the unitary QSP into two parts, one for the real part of the desired state and one for the remaining complex part, is derived, and the role of the single lambda-type constructor is discussed. The complete circuit for the $n=2$ case is shown as an example, and Section~\ref{sec:results} presents the results of the simulation test and the comparison with M\"ott\"onen et al.'s QSP~\cite{mottonen2005transformation} and with the implementation of the original version of Sun et al.'s QSP~\cite{belli2025implementation,QSP-Sun}. Finally, Section \ref{sec:conclusions} concludes the paper with a summary of the main results and a discussion of future work.

\section{SUN: traditional framework, diagonal operators and ancillary qubits}\label{sec:qsp_sun}

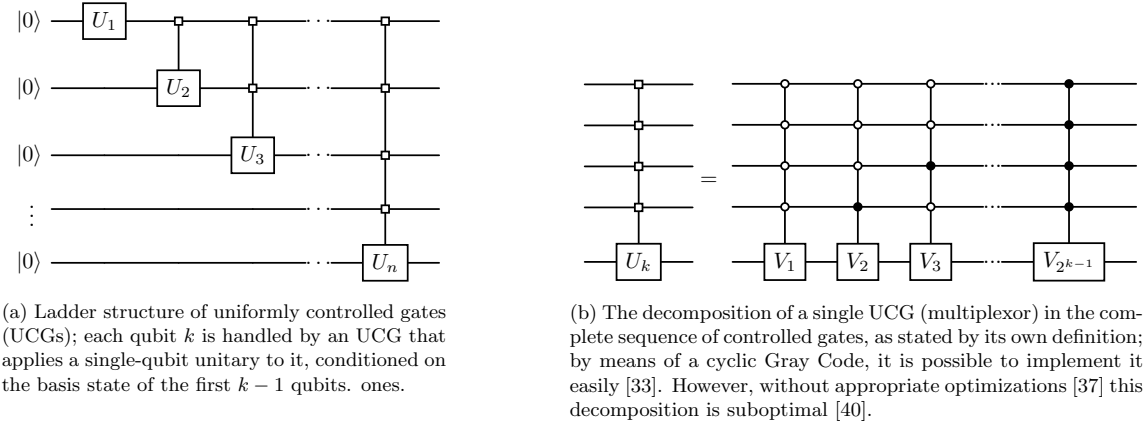
\begin{figure}[htbp]
\centering
\resizebox{0.35\textwidth}{!}{
\subcaptionbox{Ladder structure of uniformly controlled gates (UCGs); each qubit $k$ is handled by an UCG that applies a single-qubit unitary to it, conditioned on the basis state of the first $k-1$ qubits. ones.\label{fig:qsp_structure}}{
\begin{quantikz}
    \lstick{$|0\rangle$}&\gate{U_1}&\octrl[style=rectangle]{1}&\octrl[style=rectangle]{1}&\push{\ldots}&\octrl[style=rectangle]{1}&\\
    \lstick{$|0\rangle$}&&\gate{U_2}&\octrl[style=rectangle]{1}&\push{\ldots}&\octrl[style=rectangle]{1}&\\
    \lstick{$|0\rangle$}&&&\gate{U_3}&\push{\ldots}&\octrl[style=rectangle]{1}&\\
    \lstick{$\vdots\mbox{ }$}&&&&\push{\ldots}&\octrl[style=rectangle]{1}&\\
    \lstick{$|0\rangle$}&&&&\push{\ldots}&\gate{U_{n}}&
\end{quantikz}}}
\qquad\qquad
\resizebox{0.45\textwidth}{!}{
\subcaptionbox{The decomposition of a single UCG (multiplexor) in the complete sequence of controlled gates, as stated by its own definition; by means of a cyclic Gray Code, it is possible to implement it easily~\cite{mottonen2004quantum}. However, without appropriate optimizations~\cite{iten2016quantum} this decomposition is suboptimal~\cite{sun2023asymptotically}.\label{fig:multiplexor}}{
\begin{quantikz}
    &\octrl[style=rectangle]{1}&\\
    &\octrl[style=rectangle]{1}&\\
    &\octrl[style=rectangle]{1}&\\
    &\octrl[style=rectangle]{1}&\\
    &\gate{U_k}& 
\end{quantikz}=
\begin{quantikz}
    &\octrl{1}&\octrl{1}&\octrl{1}&\push{...}&\ctrl{1}&\\
    &\octrl{1}&\octrl{1}&\octrl{1}&\push{...}&\ctrl{1}&\\
    &\octrl{1}&\octrl{1}&\ctrl{1}&\push{...}&\ctrl{1}&\\
    &\octrl{1}&\ctrl{1}&\octrl{1}&\push{...}&\ctrl{1}&\\
    &\gate{V_1}&\gate{V_2}&\gate{V_3}&\push{...}&\gate{V_{2^{k-1}}}&
\end{quantikz}}}
\caption{QSP traditional framework.}
\end{figure}

In~\cite{sun2023asymptotically}, Sun et al. tightly characterized the depth-size complexity of the QSP problem, except for a logarithmic gap in the small parametric range $m=\left[\omega\left(\frac{2^n}{n\log n}\right),o(2^n)\right]$. Their algorithms, which also cover the case without ancillary qubits, allow for the construction of optimal quantum circuits where the upper and lower depth bounds coincide exactly in the two width ranges $m=\mathcal{O}\left(\frac{2^n}{n\log n}\right)$ and $m=\Omega(2^n)$. Improving on the work presented in~\cite{belli2025implementation}, regarding the first implementation of SUN, this paper focuses on the first of the two ranges with optimal \emph{asymptotic} complexity. In this specific domain, the traditional framework for QSP~\cite{mottonen2004transformation} can be used; it consists of a ladder structure of Uniformly Controlled Gates (UCGs) throughout the quantum register, as shown in Figure~\ref{fig:qsp_structure}.
In matrix form, each UCG is a block-diagonal operator $U_k$ belonging to $\mathbb{C}^{2^k\times2^k}$, where each block $V_j\in U(2)$ is a single-qubit gate for any $1\leqslant j\leqslant 2^{k-1}$, in turn decomposable as $V_j=e^{i\alpha_j}R_z(\beta_j)R_y(\gamma_j)R_z(\delta_j)$ for some parameters $(\alpha,\beta,\gamma,\delta)\in\mathbb{R}$.

\begin{figure}[htbp]
    \centering
    \resizebox{0.15\textwidth}{!}{
        \begin{quantikz}
        &\octrl[style=rectangle]{1}&\\
        &\octrl[style=rectangle]{1}&\\
        &\octrl[style=rectangle]{1}&\\
        &\gate{U_n}&
        \end{quantikz}}$\equiv$
    \resizebox{0.66\textwidth}{!}{    
        \begin{quantikz}
        &\octrl[style=rectangle]{1}&&\octrl[style=rectangle]{1}&&\octrl[style=rectangle]{1}&\gate[3]{D(e^{i\vec{\alpha}})}&\\
        &\octrl[style=rectangle]{1}&&\octrl[style=rectangle]{1}&&\octrl[style=rectangle]{1}&&\\
        &\octrl[style=rectangle]{1}&&\octrl[style=rectangle]{1}&&\octrl[style=rectangle]{1}&&\\
        &\gate{F_n(R_z(\vec{\delta}))}&\gate{HS^\dag}&\gate[]{F_n(R_z(\vec{\gamma}))}&\gate{SH}&\gate[]{F_n(R_z(\vec{\beta}))}&&\\
        \end{quantikz}}
\caption{UCG in terms of diagonal operators $F_n(R_z)$.}
\label{fig:ucg_with_diagonals}
\end{figure}
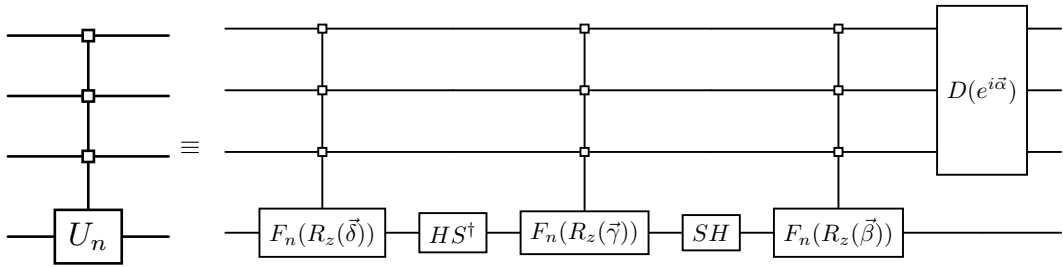

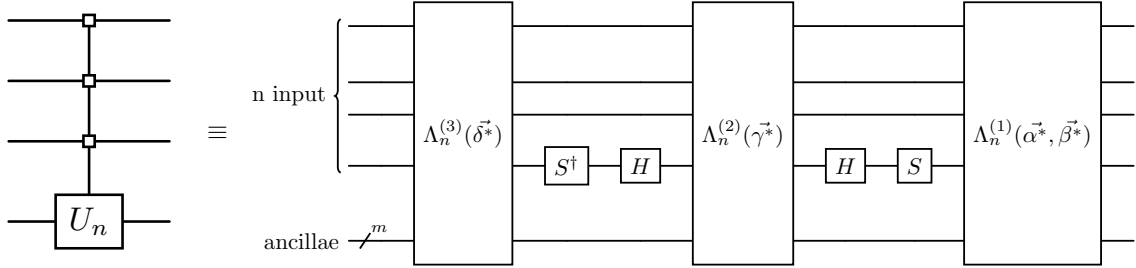
\begin{figure}[htbp]
\centering
\resizebox{0.15\textwidth}{!}{
    \begin{quantikz}
    &\octrl[style=rectangle]{1}&\\
    &\octrl[style=rectangle]{1}&\\
    &\octrl[style=rectangle]{1}&\\
    &\gate{U_n}& 
    \end{quantikz}}$\quad\equiv$   
\resizebox{0.7\textwidth}{!}{    
    \begin{quantikz}
    \lstick[4]{n input}&&\gate[5]{\Lambda_n^{(3)}(\vec{\delta^*})}&&&\gate[5]{\Lambda_n^{(2)}(\vec{\gamma^*})}&&&\gate[5]{\Lambda_n^{(1)}(\vec{\alpha^*},\vec{\beta^*})}&\\
    &&&&&&&&&\\
    &&&&&&&&&\\
    &&&\gate{S^\dag}&\gate{H}&&\gate{H}&\gate{S}&&\\
    \lstick{ancillae}&\qwbundle{m}&&&&&&&&
    \end{quantikz}}
\caption{UCG in terms of $\Lambda$-type operators.}
\label{fig:ucg_with_lambdas}
\end{figure}

Actually, it is convenient to rewrite each block $V_j$ only in terms of diagonal operators thanks to the change of gate set $R_y(\gamma_j)\equiv SH\cdot R_z(\gamma_j)\cdot HS^\dag$. In this way, the $n$-qubit extension for the arbitrary UCG $U_n$ becomes:
\begin{equation}\label{eqn:U_decomp_1}
    U_n=[\text{diag}(e^{i\alpha_1},e^{i\alpha_2},...,e^{i\alpha_{2^{n-1}}})\otimes\mathbb{I}_1]\cdot F_n[R_z(\vec{\beta})]\cdot[\mathbb{I}_{n-1}\otimes(SH)]\cdot F_n[R_z(\vec{\gamma})]\cdot [\mathbb{I}_{n-1}\otimes(HS^\dag)]\cdot F_n[R_z(\vec{\delta})]
\end{equation}
where $F_n[R_z(\vec{\theta})]=\text{diag}(R_z(\theta_1),R_z(\theta_2),...,R_z(\theta_{2^{n-1}}))$. The equivalent circuit is shown in Figure~\ref{fig:ucg_with_diagonals}, where the last two multi-qubit gates can be merged, being complex phase diagonal operators, thus decomposing the generic UCG with only three phase diagonal operators.
It is important to note that the \emph{central} block-diagonal operator $F_n(R_z(\vec{\gamma}))$ is solely responsible for preparing the real part of the desired quantum state, as the vector $\vec{\gamma}$ collects the parameters of the blocks $R_y$ after the change of the gate set. Up to global phases\footnote{A detailed mathematical proof is provided in~\cite{belli2025implementation}.}, this is already the decomposition used in~\cite{sun2023asymptotically}, where diagonal $\Lambda$-type operators,
\begin{equation}\label{eqn:lambda}
\Lambda_n(\vec{\theta})=\text{diag}(1,e^{i\theta_1},e^{i\theta_2},...,e^{i\theta_{2^n-1}})\;\in\mathbb{C}^{2^n\times2^n}\quad\mbox{for}\quad\vec{\theta}=(\theta_1,\theta_2,...,\theta_{2^{n}-1})\in\mathbb{R}^{2^{n}-1}
\end{equation}
are introduced to efficiently parallelize on an ancillary register~\cite{belli2025implementation}. Therefore, the arbitrary UCG $U_n$ can be rewritten as:
\begin{equation}\label{eqn:U_decomp_2}
        U_n=\,e^{i(\alpha_1-\beta_1)}\Lambda_n^{(1)}(\vec{\alpha^*},\vec{\beta^*})\cdot[\mathbb{I}_{n-1}\otimes(SH)]\cdot e^{-i\gamma_1}\Lambda_n^{(2)}(\vec{\gamma^*})\cdot[\mathbb{I}_{n-1}\otimes(HS^\dag)]\cdot e^{-i\delta_1}\Lambda_n^{(3)}(\vec{\delta^*})
\end{equation}
where  $\vec{\alpha^*},\vec{\beta^*},\vec{\gamma^*},\vec{\delta^*}\in\mathbb{R}^{2^n-1}$ are the vectors that group the updated $\Lambda_n$ parameters after the collection of global phases. The corresponding circuit is shown in Figure~\ref{fig:ucg_with_lambdas}.
In the end, the traditional QSP circuit of Figure~\ref{fig:qsp_structure} reduces to the concatenation of $\Lambda_n$ circuits, interspersed with the two unitary operators derived from the pair of gates $S$ and $H$.

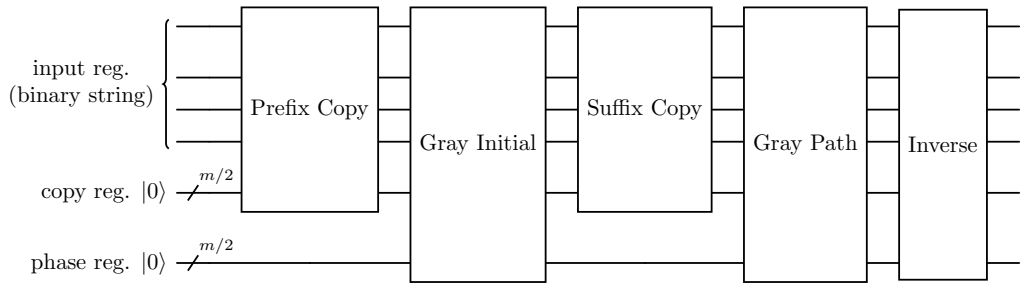
\begin{figure}[htbp]
\centering
\resizebox{0.8\textwidth}{!}{
    \begin{quantikz}
    \lstick[4]{input reg.\\(binary string)}&&\gate[5]{\mbox{Prefix Copy}}&\gate[6]{\mbox{Gray Initial}}&\gate[5]{\mbox{Suffix Copy}}&\gate[6]{\mbox{Gray Path}}&\gate[6]{\mbox{Inverse}}&\\
    &&&&&&&\\
    &&&&&&&\\
    &&&&&&&\\
    \lstick{copy reg. $|0\rangle$}&\qwbundle{m/2}&&&&&&\\
    \lstick{phase reg. $|0\rangle$}&\qwbundle{m/2}&&&&&&
    \end{quantikz}}
\caption{$\Lambda_n$ quantum circuit divided into 5 sub-unitaries (stages).}
\label{fig:lambda_circuit}
\end{figure}

\begin{figure}[htbp]
    \centering
    \includegraphics[width=1\textwidth]{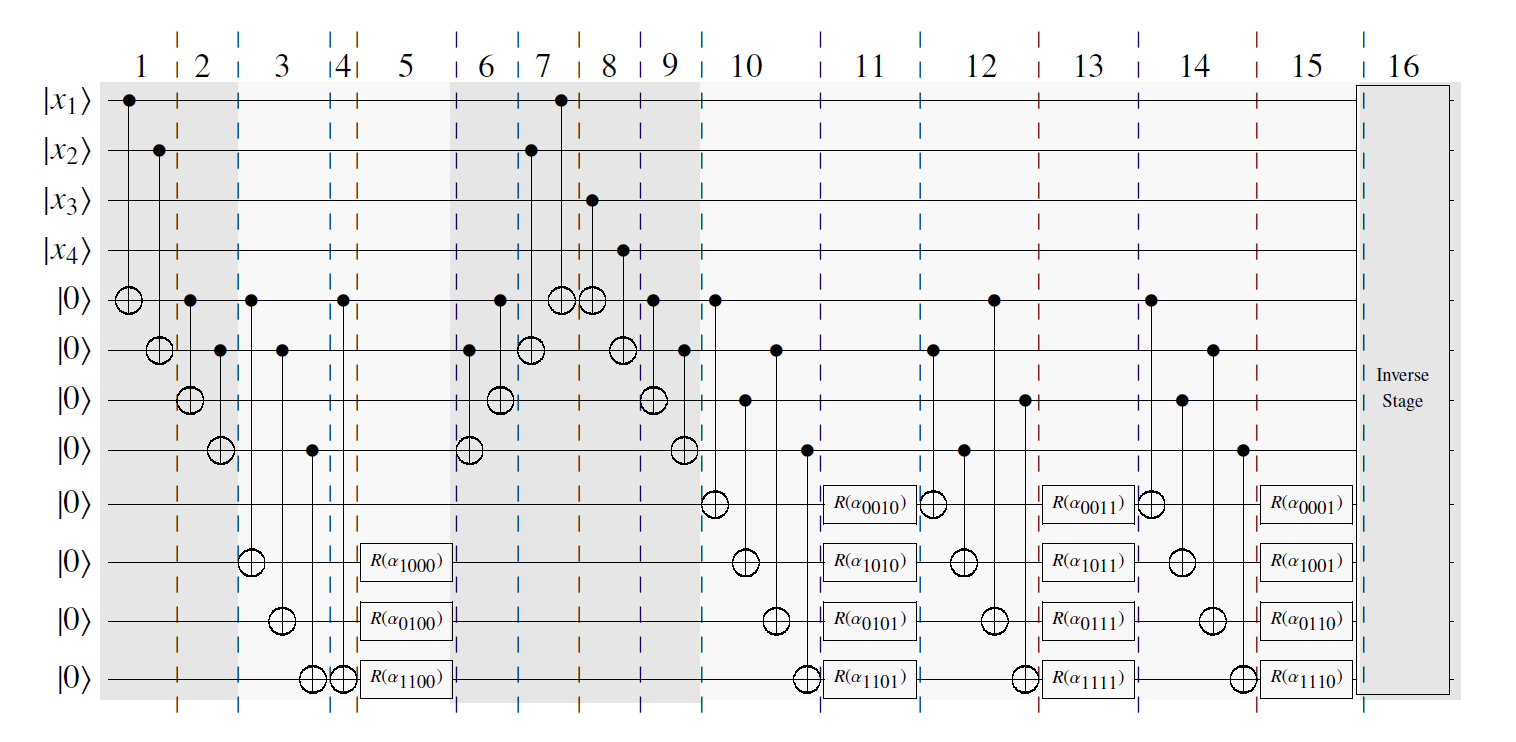}
\caption{Quantum circuit for $\Lambda_4$. Figure taken from~\cite{sun2021asymptotically}.}
\label{fig:lambda4}
\end{figure}

In~\cite[Section IV]{sun2023asymptotically}, the procedure to implement any operator $\Lambda_n\in\mathbb{C}^{2^n\times2^n}$ is discussed and proven. The equivalent quantum circuit has depth $\mathcal{O}(\log_2m +\frac{2^n}{m})$ and width $\mathcal{O}(2^n)$, within the ancillary range $m\in[2n,\frac{2^n}{n}]$. As shown in Equation~\ref{eqn:lambda}, matrices $\Lambda_n$ perform phase shifts on each state of the computational basis, a task that can be efficiently parallelized through sums in Fourier space~\cite{belli2025implementation}. The key algorithm to design the $\Lambda_n$ circuit has 5 stages that define the formalism for the scalable implementation; they are depicted from a high-level perspective in Figure~\ref{fig:lambda_circuit}. The details of their implementation, as well as the management of all parameters, are discussed in~\cite{belli2025implementation}. This work does not modify the basic implementation of operators $\Lambda_n$ in any way, and Figure~\ref{fig:lambda4} shows the case for $n=4$ as a reference example, where the 5 stages are highlighted with alternating colored bands.

\subsection{Complexity analysis for SUN}

This section retraces the crucial steps of the proof concerning the complexity of SUN, as presented in~\cite{sun2023asymptotically}, which will be relevant in Section~\ref{sec:new_decomposition} to formally define the improvement term of its optimized version, OSUN. As stated in~\cite[Lemma 4]{sun2023asymptotically}, every unitary operator $\Lambda_n\in\mathbb{C}^{2^n\times2^n}$ can be implemented by a quantum circuit of depth $\mathcal{O}(\log_2 m+\frac{2^n}{m})$, with $m\in[2n,\frac{2^n}{n}]$ ancillary qubits\footnote{This statement has been proved in~\cite[Section IV]{sun2023asymptotically}.}. Thus, it is possible to define the quantity $D_{\Lambda}(n,m)=\log_2 m+\frac{2^n}{m}$ as the representative of the asymptotic upper-bound described in~\cite[Lemma 4]{sun2023asymptotically}. In the lower bound $m=2n$ of the parametric range, corresponding to the case implemented in~\cite{belli2025implementation}, this quantity becomes $D_{\Lambda}(n,2n)=\log_2 n+1+\frac{2^{n-1}}{n}$. In the case $m=0$, the operator $\Lambda_n$ can be realized with depth $\mathcal{O}(\frac{2^n}{n})$, a result stated by~\cite[Lemma 5]{sun2023asymptotically}.

Then, considering that the QSP circuit is built upon increasing levels of UCG (Figure~\ref{fig:qsp_structure}) and that each UCG requires three $\Lambda$-type constructors, four single-gate layers, and one global phase (Eq.~\ref{eqn:U_decomp_2}), the overall QSP circuit has depth, for $m\geqslant0$:
$$
D(n,m)=\sum_{k=1}^{n}[3D_{\Lambda}(k,m)+4]+1=3\sum_{k=1}^{n}D_{\Lambda}(k,m)+4n+1
$$
where $D_{\Lambda}(k,m)$ is the depth of each $\Lambda$-type constructor at level $k$, for $k\in[1,n]$ on the entire quantum register. It is shown in~\cite[Lemma 6]{sun2023asymptotically} that each UCG $U_n$ can be implemented with a circuit of depth $O(n+\frac{2^n}{n+m})$ for $m\geqslant0$. For our purposes, it is useful to make explicit the construction of the latter single smooth bound. One can take the quantity $\tilde{D}(n,m)=\min\{\mathcal{O}(\log_2 m+\frac{2^n}{m}),\mathcal{O}(\frac{2^n}{n})\}$ for any $m\in[0,\frac{2^n}{n}]$ and show that $\tilde{D}(n,m)\subseteq\mathcal{O}(n+\frac{2^n}{m+n})$. Indeed, for $2n\leqslant m\leqslant\frac{2^n}{n}$, the two terms of $\mathcal{O}(\log_2 m+\frac{2^n}{m})$ are estimated, respectively, 
$$
\log_2 m\leqslant\log_2\left(\frac{2^n}{n}\right)=n-\log_2 n=\mathcal{O}(n)
$$
and
$$
\frac{2^n}{m}=\frac{m+n}{m}\cdot\frac{2^n}{m+n}\leqslant2\cdot\frac{2^n}{m+n}\qquad\left(\frac{m+n}{m}\leqslant2\right)
$$
proving that $\mathcal{O}(\log_2 m+\frac{2^n}{m})\subseteq\mathcal{O}(n+\frac{2^n}{n+m})$.
Instead, for $0\leqslant m<2n$, it is worth noting that $\frac{2^n}{m+n}\geqslant\frac{2^n}{2n}=\frac{1}{2}\cdot\frac{2^n}{n}$, which shows that $\mathcal{O}(\frac{2^n}{n})\subseteq\mathcal{O}(n+\frac{2^n}{n+m})$.

Therefore, according to the SUN procedure, a $n$-qubit QSP circuit with $m\geqslant0$ ancillary qubits can be designed in depth $\mathcal{O}(n^2+\frac{2^n}{n+m})$, as stated in~\cite[Lemma 7]{sun2023asymptotically}. In the range $2n\leqslant m\leqslant\frac{2^n}{n^2} (\leqslant\frac{2^n}{n})$, previous results can be combined to derive the depth upper-bound
$$
\mathcal{O}\left[\sum_{j=1}^{n}(\log_2 m+\frac{2^j}{m})+n\right]=\mathcal{O}\left[(\log_2 m)n+\frac{1}{m}\sum_{j=1}^{n}2^j+n\right]=\mathcal{O}\left(n^2+\frac{2^n}{m}\right)
$$
while for $m=0$, the asymptotic complexity is $\mathcal{O}\left[\sum_{j=1}^{n}\frac{2^j}{j}+n\right]=\mathcal{O}(\frac{2^n}{n})$.

\section{A new decomposition with a single $\Lambda$-operator for each uniformly controlled gate}\label{sec:new_decomposition}

As proved in~\cite{belli2025srbb}, the traditional structure for quantum state preparation can be split into two unitaries: one devoted to the preparation of the real part of the desired state and the other to the complex component. This separation is allowed by the conventional choice of initializing the state preparation circuits with $|0\rangle^{\otimes n}$, which in turn implies that only the first column of the total QSP unitary is relevant for preparation purposes. In this way, by concatenating the circuit block $U_{mod}$ responsible for the preparation of the real part with a diagonal unitary $D_{ph}$ of complex phases, a complete QSP circuit is obtained, as illustrated in Figure~\ref{fig:qsp_structure_reduced}.
\begin{figure}[htbp]
\centering
    \begin{quantikz}
    \lstick{$|0\rangle^{\otimes n}$}&\gate[1]{U_{mod}}&\gate[1]{D_{ph}}&\rstick{$|\psi\rangle=\sum_{i=0}^{2^n-1}c_i|i\rangle$}
    \end{quantikz}
\caption{QSP structure with the separation of the real part from the complex one.}
\label{fig:qsp_structure_reduced}
\end{figure}
The unitary $U_{mod}$ can be implemented through the traditional QSP framework~\cite{mottonen2004transformation}, which is composed of Uniformly Controlled Gates (UCGs) arranged in a ladder across the entire quantum register; but, since it only needs to prepare the real part of the desired state, each UCG is a $R_y$-block diagonal matrix:
\begin{equation}
V_j=\cancel{e^{i\alpha_j}}\cancel{R_z(\beta_j)}R_y(\gamma_j)\cancel{R_z(\delta_j)}\in SU(2)
\end{equation}
In order to effectively parallelize according to SUN, gate set switching is mandatory.
\begin{equation}
R_y(\gamma_j)\equiv SH\cdot R_z(\gamma_j)\cdot HS^\dag
\end{equation}
where now, contrary to the decomposition of Equation (\ref{eqn:U_decomp_2}), each UCG $U_n$ is encoded in a single matrix $\Lambda_n^{(2)}$, as illustrated in Figure~\ref{fig:ucg_reduced}. For this reason, hereafter this algorithm will be called \emph{single-$\Lambda$ ancillae-based QSP}. In analogy to Figure~\ref{fig:ucg_with_diagonals}, the circuit corresponding to the unitary $U_{mod}$, which exploits the parallelization on the ancillary register, is represented in Figure~\ref{fig:ucg_with_lambda2}.
\begin{figure}[htbp]
\centering
    \begin{quantikz}
    &\octrl[style=rectangle]{1}&\\
    &\octrl[style=rectangle]{1}&\\
    &\octrl[style=rectangle]{1}&\\
    &\gate{U_n}& 
    \end{quantikz}=
    \begin{quantikz}
    &&\octrl[style=rectangle]{1}&&\\
    &&\octrl[style=rectangle]{1}&&\\
    &&\octrl[style=rectangle]{1}&&\\
    &\gate{HS^\dag}&\gate{F_n(R_z(\vec{\gamma}))}&\gate{SH}&
    \end{quantikz}
\caption{Reduced quantum circuit for the UCG.}
\label{fig:ucg_reduced}
\end{figure}
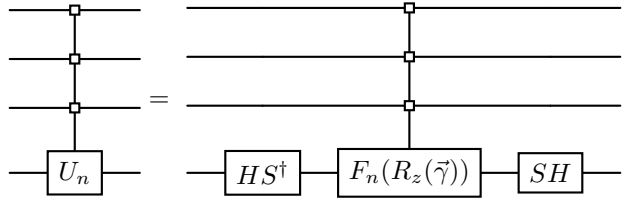
\begin{figure}[htbp]
\centering
\resizebox{0.15\textwidth}{!}{
    \begin{quantikz}
    &\octrl[style=rectangle]{1}&\\
    &\octrl[style=rectangle]{1}&\\
    &\octrl[style=rectangle]{1}&\\
    &\gate{U_n}& 
    \end{quantikz}}$\quad\equiv$   
\resizebox{0.45\textwidth}{!}{    
    \begin{quantikz}
    \lstick[4]{n input}&&&\gate[5]{\Lambda_n^{(2)}(\vec{\gamma^*})}&&&\\
    &&&&&&\\
    &&&&&&\\
    &\gate{S^\dag}&\gate{H}&&\gate{H}&\gate{S}&\\
    \lstick{ancillae}&\qwbundle{m}&&&&&
    \end{quantikz}}
\caption{UCG in terms of a single $\Lambda$-type operators.}
\label{fig:ucg_with_lambda2}
\end{figure}
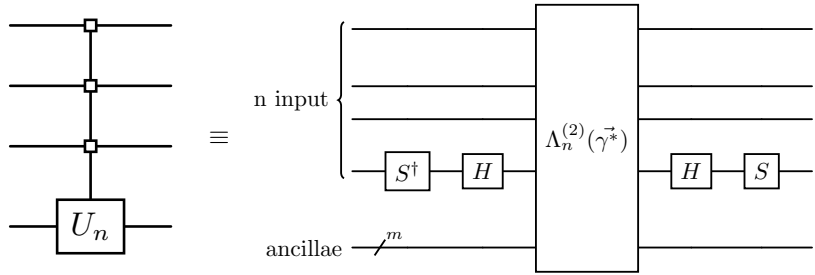
In~\cite[Section 2]{belli2025implementation} it is shown that the ladder structure of UCG induces a simple recursive scheme on the first column of the total unitary, useful for deriving the parameters of the $R_y$ gates according to the traditional Binary Search Tree (BST) method.

On the other hand, the diagonal unitary $D_{ph}$ can be implemented directly via a $\Lambda$-type operator after an appropriate global phase recollection. If $|\psi\rangle=\sum_{k=0}^{2^n-1}|c_k|e^{i\phi_k}|k\rangle$, the complex phases $e^{i(\phi_k-\phi_0)}$ will be the diagonal entries of $D_{ph}$ for $0\leqslant k\leqslant 2^n-1$, where $e^{i\phi_0}$ is the global phase. 

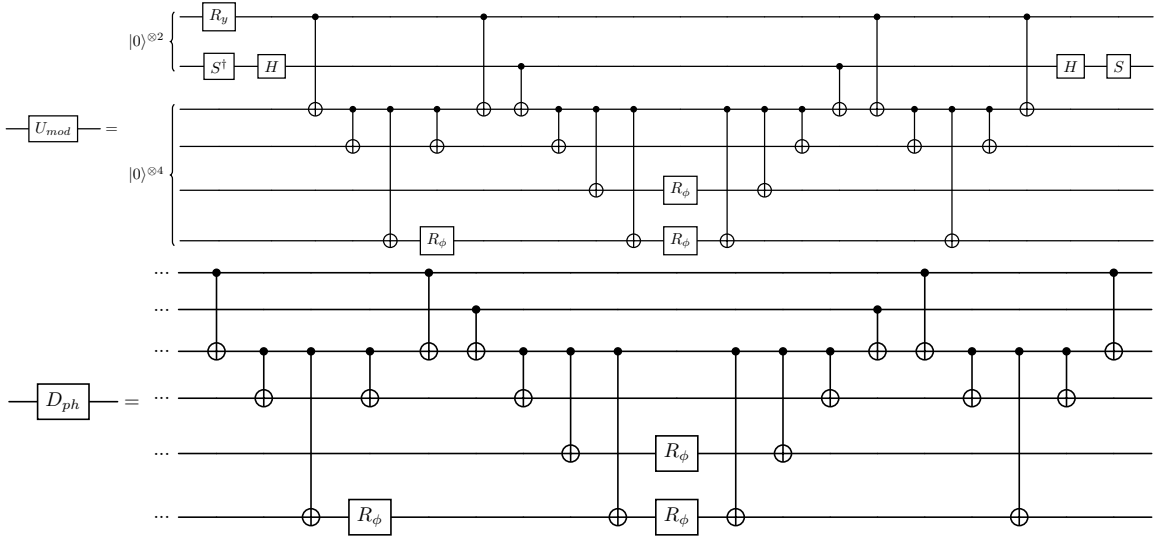
\begin{figure}[htbp]
    \centering
    \resizebox{0.9\textwidth}{!}{
        \centering
        \begin{quantikz}
        &\gate{U_{mod}}& 
        \end{quantikz}=
        \begin{quantikz}
        \lstick[2]{$|0\rangle^{\otimes2}$}&\gate{R_y}&&\ctrl{2}&&&&\ctrl{2}&&&&&&&&&&\ctrl{2}&&&&\ctrl{2}&&&\\
        &\gate{S^\dag}&\gate{H}&&&&&&\ctrl{1}&&&&&&&&\ctrl{1}&&&&&&\gate{H}&\gate{S}&\\
        \lstick[4]{$|0\rangle^{\otimes4}$}&&&\targ{}&\ctrl{1}&\ctrl{3}&\ctrl{1}&\targ{}&\targ{&}&\ctrl{1}&\ctrl{2}&\ctrl{3}&&\ctrl{3}&\ctrl{2}&\ctrl{1}&\targ{}&\targ{}&\ctrl{1}&\ctrl{3}&\ctrl{1}&\targ{}&&&\\
        &&&&\targ{}&&\targ{}&&&\targ{}&&&&&&\targ{}&&&\targ{}&&\targ{}&&&&\\
        &&&&&&&&&&\targ{}&&\gate{R_{\phi}}&&\targ{}&&&&&&&&&&\\
        &&&&&\targ{}&\gate{R_{\phi}}&&&&&\targ{}&\gate{R_{\phi}}&\targ{}&&&&&&\targ{}&&&&&
        \end{quantikz}}
    \resizebox{0.9\textwidth}{!}{
        \centering
        \begin{quantikz}
        &\gate{D_{ph}}& 
        \end{quantikz}=
        \begin{quantikz}
        \lstick{...}&\ctrl{2}&&&&\ctrl{2}&&&&&&&&&&\ctrl{2}&&&&\ctrl{2}&\\
        \lstick{...}&&&&&&\ctrl{1}&&&&&&&&\ctrl{1}&&&&&&\\
        \lstick{...}&\targ{}&\ctrl{1}&\ctrl{3}&\ctrl{1}&\targ{}&\targ{}&\ctrl{1}&\ctrl{2}&\ctrl{3}&&\ctrl{3}&\ctrl{2}&\ctrl{1}&\targ{}&\targ{}&\ctrl{1}&\ctrl{3}&\ctrl{1}&\targ{}&\\
        \lstick{...}&&\targ{}&&\targ{}&&&\targ{}&&&&&&\targ{}&&&\targ{}&&\targ{}&&\\
        \lstick{...}&&&&&&&&\targ{}&&\gate{R_{\phi}}&&\targ{}&&&&&&&&\\
        \lstick{...}&&&\targ{}&\gate{R_{\phi}}&&&&&\targ{}&\gate{R_{\phi}}&\targ{}&&&&&&\targ{}&&&
        \end{quantikz}}
        
\caption{QSP circuit for $n=2$ divided into the two unitary blocks $U_{mod}$ and $D_{ph}$.}
\label{fig:qsp_total-n2}
\end{figure}

Ultimately, according to the new decomposition summarized in the 2-unitary block structure of Figure~\ref{fig:qsp_structure_reduced}, the overall reduced-depth quantum circuit for the preparation of a 2-qubit (4-ancilla) arbitrary state is depicted in Figure~\ref{fig:qsp_total-n2}.

\subsection{Complexity analysis for OSUN}

The optimized version of SUN described in Section~\ref{sec:new_decomposition}, hereafter referred to as OSUN, uses only one (instead of three) circuit block $\Lambda_k$ for each UCG dedicated to the real part of the desired state, and a final block $\Lambda_n$ for the imaginary part, in the parametric range $2n\leqslant m\leqslant\frac{2^n}{n}$. Therefore, the whole $n$-qubit QSP circuit has depth $D(n,m)=\sum_{k=1}^{n}[D_{\Lambda}(k,m)+4]+1+D_{\Lambda}(n,m)$, which essentially leads to the same complexity: the improvement term is hidden in the constant factors of the two main contributions. For this reason, reproducing the complexity analysis of Section~\ref{sec:qsp_sun} using a model with \emph{explicit constants} will allow us to formally define the improvement.

\subsubsection*{SUN complexity model with explicit constant}
In the parametric range $2n\leqslant m\leqslant\frac{2^n}{n}$, each operator $\Lambda_j$ has depth $D_{\Lambda}^{anc}(j,m)\leqslant a\log_2 m+b\frac{2^j}{m}$ for any real numbers $a,b>0$ and $j\in[1,n]$. In each level $j$, the UCG $U_j$ also counts four single-qubit layers and one global phase, for a total cost indicated below as $c$. When no ancillary qubits are available, each operator $\Lambda_j$ has depth $D_{\Lambda}^{no-anc}(j)\leqslant d\frac{2^j}{j}$, for any real $d>0$. Therefore, the construction of a single smooth bound $m\geqslant0$ for the depth of each UCG $U_j$ leads one to consider:
$$
D_j^{case}(m)\leqslant3D_{\Lambda}^{case}(j,m)+c=\left\{
\begin{array}{ll}
     3a\log_2 m +3b\frac{2^j}{m}+c&\mbox{case}=anc\\
     3d\frac{2^j}{j}+c&\mbox{case}=no-anc 
\end{array}
\right.
$$
In the following, it is shown that there exists a real $k>0$ such that $D_j^{case}(m)\leqslant k(n+\frac{2^n}{n+m})$ for $m\geqslant0$, considering two cases:  
\begin{itemize}
    \item[a)]in the range $2n\leqslant m\leqslant\frac{2^n}{n}$, $\log_2 m\leqslant\log_2(\frac{2^n}{n})=n-\log_2 n\leqslant n$, thus $3a\log_2 m\leqslant3an$. Instead, $\frac{2^j}{m}=\frac{m+n}{m}\cdot\frac{2^j}{m+n}\leqslant\frac{3}{2}\frac{2^j}{m+n}$, thus $3b\frac{2^j}{m}\leqslant\frac{9}{2}b\frac{2^j}{m+n}$. Therefore,
    $$
    D_j^{anc}(m)\leqslant3an+\frac{9b}{2}\frac{2^j}{m+n}+c\leqslant k_1\left(n+\frac{2^n}{m+n}\right)$$
    for $k_1\geqslant\max\{3a,\frac{9b}{2},c\}$.
    \item[b)]in the range $0\leqslant m<2n$, it is simple to see that $\frac{2^j}{j+m}\in(\frac{2^j}{2n+j},\frac{2^j}{j}]$, since $j\leqslant j+m<2n+j$. Then $\frac{2^j}{j}\leqslant(\frac{2n}{j}+1)\frac{2^j}{j+m}$ and
    $$
    D_j^{no-anc}(m)\leqslant 3d\frac{2^j}{j}+c\leqslant 3d\left(\frac{2n}{j}+1\right)\frac{2^j}{j+m}+c\leqslant 3d\left(\frac{2n}{j}+1\right)\left(n+\frac{2^n}{n+m}\right)+c\leqslant k_2\left(n+\frac{2^n}{n+m}\right)
    $$
    for $k_2\geqslant\max\left\{3d\left(\frac{2n}{j}+1\right),c\right\}$.
\end{itemize}
As a consequence, taking $k=\max\{k_1,k_2\}$, $D_j^{case}(m)\in\mathcal{O}(n+\frac{2^n}{n+m})$. 

The total depth of the circuit can be computed, respectively, in the range $2n\leqslant m\leqslant\frac{2^n}{n}$ by
\begin{equation}\label{eqn:D_anc}
D(n,m)^{anc}=\sum_{j=1}^{n}D_j^{anc}(m)\leqslant3an\log m+\frac{3b}{m}(2^{n+1}-2)+cn=\mathcal{O}(n^2)+\mathcal{O}\left(\frac{2^n}{m}\right)=\mathcal{O}\left(n^2+\frac{2^n}{m}\right)
\end{equation}
and in the range $0\leqslant m<2n$ by
$$
D(n)^{no-anc}=\sum_{j=1}^{n}D_j^{no-anc}\leqslant3dS_n+cn
$$
where $S_n:=\sum_{j=1}^{n}\frac{2^j}{j}=\sum_{j=1}^{\lfloor n/2\rfloor}\frac{2^j}{j}+\sum_{\lfloor n/2\rfloor+1}^{n}\frac{2^j}{j}$. In the latter expression,
$$
\sum_{j=1}^{\lfloor n/2\rfloor}\frac{2^j}{j}\leqslant\sum_{j=1}^{\lfloor n/2\rfloor}2^j\leqslant2^{n/2+1}
$$
while
$$
\sum_{\lfloor n/2\rfloor+1}^{n}\frac{2^j}{j}\leqslant\frac{2}{n}\sum_{\lfloor n/2\rfloor+1}^{n}2^j\leqslant\frac{2}{n}2^{n+1}
$$
so that $S_n\leqslant2^{n/2+1}+\frac{4}{n}2^n\leqslant k_3\frac{2^n}{n}$ and
\begin{equation}\label{eqn:D_noanc}
D(n)^{no-anc}\leqslant3dk_3\frac{2^n}{n}+cn=\mathcal{O}\left(\frac{2^n}{n}\right)
\end{equation}
Consistency at the edge $m=2n$ is ensured by
$$
D(n,2n)^{anc}=3an\log(2n)+\frac{3b}{2n}(2^{n+1}-2)+cn\approx 3an\log n+\frac{3b}{n}2^n+\mathcal{O}(n)=\mathcal{O}\left(\frac{2^n}{n}\right)
$$
meaning that the two regimes connect without any jumps.

\subsubsection*{OSUN complexity model with explicit constant}

Based on the algebraic reduction described in Section~\ref{sec:new_decomposition}, valid in the parametric range $m\in[2n,\frac{2^n}{n}]$, the optimized version OSUN has a ladder structure of $R_y$-based UCGs starting from level $j=2$. This is because level $j=1$ only has one uncontrolled gate $R_y$ (refer to Figure~\ref{fig:qsp_total-n2} as an example), which is possible due to the hypothesis of preparing the real part of the desired state separately from the complex part. This simplification has the following consequences:
\begin{itemize}
    \item level $j=1$ has a constant depth $c_0=2$;
    \item for each level $j\in[2,n]$, there are two single-gate layers, which we denote as cost $c_1=2$;
    \item for each level $j\in[2,n]$, there is one $\Lambda$-type constructor for the real part, plus another one for $j=n$ regarding the imaginary part of the desired state;
    \item $c_3$ is the cost of the global phase.
\end{itemize}
Then, based on the new decomposition summarized by Figure~\ref{fig:ucg_with_lambda2}, the total depth can be evaluated as
$$
D_{opt}^{anc}(n,m)=c_0+\sum_{j=2}^{n}[D_{\Lambda}^{anc}(j,m)+c_1]+D_{\Lambda}^{anc}(n,m)+c_3
$$
where $D_{\Lambda}^{anc}(j,m)\leqslant a\log m+b\frac{2^j}{m}$ for real $a,b>0$, as before. Considering that $\sum_{j=2}^{n}b\frac{2^j}{m}=\frac{b}{m}(2^{n+1}-4)$,
\begin{equation}\label{eqn:D_anc_opt}
D_{opt}^{anc}(n,m)\leqslant an\log m+\frac{b}{m}(3\cdot2^n-4)+c_1n+\mathcal{O}(1) 
\end{equation}
From a quick comparison with Equation~\ref{eqn:D_anc}, it is clear that term $n\log m$ shows a factor of 3 improvement, while term $\frac{2^n}{m}$ shows a factor of 2 improvement; on the contrary, the linear terms are similar and absorbed by $\mathcal{O}(n^2)$ in the global bound. It is also clear that the algebraic reduction has not changed the complexity class: on the range $2n\leqslant m\leqslant\frac{2^n}{n}$, $an\log m\leqslant an^2$, so that $D_{opt}^{anc}(n,m)\in\mathcal{O}(n^2+\frac{2^n}{m})$. In conclusion, the new construction has the same asymptotic complexity but systematically lowers the prefactors of the two dominant contributions (linear-logarithmic and exponential).

For completeness, the analysis of OSUN in the range $0\leqslant m<2n$ is reported below. Applying the same assumptions as in the previous range, the total depth is given by
$$
D_{opt}^{no-anc}(n)=c_0+\sum_{j=2}^{n}[D_{\Lambda}^{no-anc}(j,m)+c_1]+D_{\Lambda}^{no-anc}(n,m)+c_3
$$
where now $D_{\Lambda}^{no-anc}(j)\leqslant d\frac{2^j}{j}$, for real $d>0$. Remembering that $S_n:=\sum_{j=1}^{n}\frac{2^j}{j}$, it is worth that $\sum_{j=2}^{n}\frac{2^j}{j}\leqslant S_n\leqslant k\frac{2^n}{n}$, so that
\begin{equation}\label{eqn:D_noanc_opt}
D_{opt}^{no-anc}(n)\leqslant d(k+1)\frac{2^n}{n}+cn+\mathcal{O}(1)=\mathcal{O}\left(\frac{2^n}{n}\right)
\end{equation}
The comparison with Equation~\ref{eqn:D_noanc} also shows, in this case, the reduction of the prefactor of the dominant term.

\section{Simulations and Results}\label{sec:results}

The \emph{single-$\Lambda$ ancillae-based} QSP algorithm has been tested using the PennyLane simulator~\cite{bergholm2018pennylane} on a Linux machine equipped with an AMD EPYC 7282 CPU and 256 GB of RAM. First, the preparation of some quantum states of interest for research was verified: the four Bell states for 2 qubits, as well as the GHZ, W, and Dicke states for 3 and 4 qubits. The results are reported in Table~\ref{tab:results_specific}, where the columns ``Time Classical'' and ``Time Quantum'' indicate, respectively, the time for determining the parameters (classical subroutine) and the time for the execution of the quantum circuit in a simulated environment. Depending on their amplitudes, not all states require the general circuit of the complex case (see Figure~\ref{fig:qsp_structure_reduced} for the general structure and Figure~\ref{fig:qsp_total-n2} as an example for 2 qubits): states with positive real coefficients can be prepared with a reduced circuit, which corresponds to the only $U_{mod}$ unitary of Figure~\ref{fig:qsp_structure_reduced}.
\begin{table}[htbp]
\centering
\resizebox{0.9\textwidth}{!}{
    \begin{tabular}{|>{\centering\arraybackslash}m{1cm}|>{\centering\arraybackslash}m{2cm}|>{\centering\arraybackslash}m{2.4cm}|>{\centering\arraybackslash}m{2.4cm}|>{\centering\arraybackslash}m{1.1cm}|>{\centering\arraybackslash}m{1.1cm}|>{\centering\arraybackslash}m{1.1cm}|>{\centering\arraybackslash}m{1.7cm}|>{\centering\arraybackslash}m{1.7cm}|}
     \hline
     \multicolumn{9}{|c|}{\textbf{Simulation results for specific quantum states}}\\
     \hline
     \hline
     $n$  & Target State & Avg Fidelity $\sigma^2\leqslant10^{-32}$ & Avg Trace Distance $\sigma^2=0.0$ & Depth & Total gates & CNOT & Avg Time Classical (s) & Avg Time Quantum (s) \\
     \hline
     \rule[-1mm]{0mm}{0.5cm}\multirow{4}*{2} & Bell $\Phi_+$ & $1-2\times10^{-16}$ & $2.78\times10^{-16}$ & 20 & 26 & 18 & $1.41\times10^{-3}$ & $3.54\times10^{-2}$ \\
     \cline{2-9}
     \rule[-1mm]{0mm}{0.5cm}& Bell $\Phi_-$ & $1-2\times10^{-16}$ & $2.78\times10^{-16}$ & 39 & 47 & 36 & $7.11\times10^{-3}$ & $4.20\times10^{-2}$ \\
     \cline{2-9}
     \rule[-1mm]{0mm}{0.5cm}& Bell $\Psi_+$ & $1-2\times10^{-16}$ & $2.78\times10^{-16}$ & 20 & 26 & 18 & $1.17\times10^{-3}$ & $3.77\times10^{-2}$ \\
     \cline{2-9}
     \rule[-1mm]{0mm}{0.5cm}& Bell $\Psi_-$ & $1-2\times10^{-16}$ & $2.78\times10^{-16}$ & 39 & 47 & 36 & $6.89\times10^{-3}$ & $2.75\times10^{-2}$ \\
     \cline{1-9}
     \rule[-1mm]{0mm}{0.5cm}\multirow{3}*{3} & GHZ & $1-6\times10^{-16}$ & $4.16\times10^{-16}$ & 43 & 67 & 48 & $4.13\times10^{-3}$ & $4.26\times10^{-2}$ \\
     \cline{2-9}
     \rule[-1mm]{0mm}{0.5cm}& W & $1-6\times10^{-16}$ & $3.05\times10^{-16}$ & 43 & 67 & 48 & $2.81\times10^{-3}$ & $4.84\times10^{-2}$ \\
     \cline{2-9}
     \rule[-1mm]{0mm}{0.5cm}& Dicke & $1-6\times10^{-16}$ & $3.05\times10^{-16}$ & 43 & 67 & 48 & $2.89\times10^{-3}$ & $5.81\times10^{-2}$ \\
     \cline{1-9}
     \rule[-1mm]{0mm}{0.5cm}\multirow{3}*{4} & GHZ & $1-9\times10^{-16}$ & $7.49\times10^{-16}$ & 74 & 142 & 104 & $8.65\times10^{-3}$ & $7.30\times10^{-2}$ \\
     \cline{2-9}
     \rule[-1mm]{0mm}{0.5cm}& W & $1-9\times10^{-16}$ & $5.83\times10^{-16}$ & 74 & 142 & 104 & $6.70\times10^{-3}$ & $8.27\times10^{-2}$ \\
     \cline{2-9}
     \rule[-1mm]{0mm}{0.5cm}& Dicke & $1-9\times10^{-16}$ & $5.27\times10^{-16}$ & 74 & 142 & 104 & $6.74\times10^{-3}$ & $7.60\times10^{-2}$ \\
     \hline
     \end{tabular}}
\caption{Preparation of well-known quantum states. The results refer to a sample of 20 states.}
\label{tab:results_specific}
\end{table}

In a second step, the effectiveness of the algorithm has been verified in the preparation of random quantum states, both dense and sparse, with real positive, real negative, or complex coefficients, from 2 to 5 qubits. The results are reported in Tables~\ref{tab:results_random_complex}, \ref{tab:results_random_positive} and \ref{tab:results_random_negative}.
\begin{table}[htbp]
\centering
\resizebox{0.9\textwidth}{!}{
    \begin{tabular}{|>{\centering\arraybackslash}m{0.8cm}|>{\centering\arraybackslash}m{1.5cm}|>{\centering\arraybackslash}m{3.1cm}|>{\centering\arraybackslash}m{3.1cm}|>{\centering\arraybackslash}m{1.2cm}|>{\centering\arraybackslash}m{1.2cm}|>{\centering\arraybackslash}m{1.7cm}|>{\centering\arraybackslash}m{1.7cm}|}
     \hline
     \multicolumn{8}{|c|}{\textbf{Simulation results for random states with complex amplitudes}}\\
     \hline
     \hline
     $n$ & Type of state & Avg Fidelity $\sigma^2\in[10^{-32},10^{-31}]$ & Avg Trace Distance $\sigma^2\in[10^{-33},10^{-32}]$ & Depth & CNOT gates & Avg Time Classical (s) & Avg Time Quantum (s) \\
     \hline
     \rule[-1mm]{0mm}{0.5cm}\multirow{2}*{2}& dense & $1-2\times10^{-16}$ & $2.43\times10^{-16}$ & \multirow{2}*{39} & \multirow{2}*{36} & $7.88\times10^{-3}$ & $6.11\times10^{-2}$ \\
     \cline{2-4}\cline{7-8}
     \rule[-1mm]{0mm}{0.5cm}& sparse & $1-3\times10^{-16}$ & $2.58\times10^{-16}$ &  &  & $1.06\times10^{-2}$ & $3.87\times10^{-2}$ \\
     \cline{1-8}
     \rule[-1mm]{0mm}{0.5cm}\multirow{2}*{3}& dense & $1-6\times10^{-16}$ & $3.79\times10^{-16}$ & \multirow{2}*{66} & \multirow{2}*{78} & $7.21\times10^{-3}$ & $8.90\times10^{-2}$ \\
     \cline{2-4}\cline{7-8}
     \rule[-1mm]{0mm}{0.5cm}& sparse & $1-3\times10^{-16}$ & $3.07\times10^{-16}$ &  &  & $3.69\times10^{-3}$ & $5.92\times10^{-2}$ \\
     \cline{1-8}
     \rule[-1mm]{0mm}{0.5cm}\multirow{2}*{4}& dense & $1-9\times10^{-16}$ & $5.71\times10^{-16}$ & \multirow{2}*{105} & \multirow{2}*{160} & $8.46\times10^{-3}$ & $8.91\times10^{-2}$ \\
     \cline{2-4}\cline{7-8}
     \rule[-1mm]{0mm}{0.5cm}& sparse & $1-9\times10^{-16}$ & $5.48\times10^{-16}$ &  &  & $8.56\times10^{-3}$ & $8.64\times10^{-2}$ \\
     \hline
     \rule[-1mm]{0mm}{0.5cm}\multirow{2}*{5}& dense & $1-2\times10^{-15}$ & $7.48\times10^{-16}$ & \multirow{2}*{166} & \multirow{2}*{276} & $6.30\times10^{-2}$ & $1.55\times10^{-1}$ \\
     \cline{2-4}\cline{7-8}
     \rule[-1mm]{0mm}{0.5cm}& sparse & $1-2\times10^{-15}$ & $7.47\times10^{-16}$ &  &  & $5.05\times10^{-2}$ & $1.39\times10^{-1}$ \\
     \hline
    \end{tabular}}
\caption{Preparation of random quantum states with complex coefficients. The results refer to a sample of 20 states.}
\label{tab:results_random_complex}
\end{table}
\begin{table}[htbp]
\centering
\resizebox{0.9\textwidth}{!}{
    \begin{tabular}{|>{\centering\arraybackslash}m{0.8cm}|>{\centering\arraybackslash}m{1.5cm}|>{\centering\arraybackslash}m{3.1cm}|>{\centering\arraybackslash}m{3.1cm}|>{\centering\arraybackslash}m{1.2cm}|>{\centering\arraybackslash}m{1.2cm}|>{\centering\arraybackslash}m{1.7cm}|>{\centering\arraybackslash}m{1.7cm}|}
     \hline
     \multicolumn{8}{|c|}{\textbf{Simulation results for random states with real positive amplitudes}}\\
     \hline
     \hline
     $n$ & Type of state & Avg Fidelity $\sigma^2\in[10^{-32},10^{-31}]$ & Avg Trace Distance $\sigma^2\in[10^{-33},10^{-30}]$ & Depth & CNOT gates & Avg Time Classical (s) & Avg Time Quantum (s) \\
     \hline
     \rule[-1mm]{0mm}{0.5cm}\multirow{2}*{2}& dense & $1-3\times10^{-16}$ & $2.61\times10^{-16}$ & \multirow{2}*{20} & \multirow{2}*{18} & $1.17\times10^{-3}$ & $5.47\times10^{-2}$ \\
     \cline{2-4}\cline{7-8}
     \rule[-1mm]{0mm}{0.5cm}& sparse & $1-3\times10^{-16}$ & $2.23\times10^{-16}$ &  &  & $1.13\times10^{-3}$ & $5.74\times10^{-2}$ \\
     \cline{1-8}
     \rule[-1mm]{0mm}{0.5cm}\multirow{2}*{3}& dense & $1-6\times10^{-16}$ & $3.58\times10^{-16}$ & \multirow{2}*{43} & \multirow{2}*{48} & $2.50\times10^{-3}$ & $6.19\times10^{-2}$ \\
     \cline{2-4}\cline{7-8}
     \rule[-1mm]{0mm}{0.5cm}& sparse & $1-6\times10^{-16}$ & $5.26\times10^{-16}$ &  &  & $2.37\times10^{-3}$ & $7.27\times10^{-2}$ \\
     \cline{1-8}
     \rule[-1mm]{0mm}{0.5cm}\multirow{2}*{4}& dense & $1-7\times10^{-16}$ & $8.98\times10^{-16}$ & \multirow{2}*{74} & \multirow{2}*{104} & $6.65\times10^{-3}$ & $9.13\times10^{-2}$ \\
     \cline{2-4}\cline{7-8}
     \rule[-1mm]{0mm}{0.5cm}& sparse & $1-7\times10^{-16}$ & $5.33\times10^{-16}$ &  &  & $6.62\times10^{-3}$ & $8.76\times10^{-2}$ \\
     \hline
     \rule[-1mm]{0mm}{0.5cm}\multirow{2}*{5}& dense & $1-2\times10^{-15}$ & $7.96\times10^{-16}$ & \multirow{2}*{120} & \multirow{2}*{190} & $6.24\times10^{-2}$ & $1.47\times10^{-1}$ \\
     \cline{2-4}\cline{7-8}
     \rule[-1mm]{0mm}{0.5cm}& sparse & $1-2\times10^{-15}$ & $6.16\times10^{-16}$ &  &  & $7.71\times10^{-2}$ & $1.32\times10^{-1}$ \\
     \hline
    \end{tabular}}
\caption{Preparation of random quantum states with real positive coefficients. The results refer to a sample of 20 states.}
\label{tab:results_random_positive}
\end{table}
\begin{table}[htbp]
\centering
\resizebox{0.9\textwidth}{!}{
    \begin{tabular}{|>{\centering\arraybackslash}m{0.8cm}|>{\centering\arraybackslash}m{1.5cm}|>{\centering\arraybackslash}m{3.1cm}|>{\centering\arraybackslash}m{3.1cm}|>{\centering\arraybackslash}m{1.2cm}|>{\centering\arraybackslash}m{1.2cm}|>{\centering\arraybackslash}m{1.7cm}|>{\centering\arraybackslash}m{1.7cm}|}
     \hline
     \multicolumn{8}{|c|}{\textbf{Simulation results for random states with real negative amplitudes}}\\
     \hline
     \hline
     $n$ & Type of state & Avg Fidelity $\sigma^2\in[10^{-32},10^{-31}]$ & Avg Trace Distance $\sigma^2\in[10^{-33},10^{-29}]$ & Depth & CNOT gates & Avg Time Classical (s) & Avg Time Quantum (s) \\
     \hline
     \rule[-1mm]{0mm}{0.5cm}\multirow{2}*{2}& dense & $1-2\times10^{-16}$ & $2.17\times10^{-16}$ & \multirow{2}*{39} & \multirow{2}*{36} & $6.37\times10^{-3}$ & $3.14\times10^{-2}$ \\
     \cline{2-4}\cline{7-8}
     \rule[-1mm]{0mm}{0.5cm}& sparse & $1-2\times10^{-16}$ & $2.64\times10^{-16}$ &  &  & $8.82\times10^{-3}$ & $3.22\times10^{-2}$ \\
     \cline{1-8}
     \rule[-1mm]{0mm}{0.5cm}\multirow{2}*{3}& dense & $1-7\times10^{-16}$ & $3.74\times10^{-16}$ & \multirow{2}*{66} & \multirow{2}*{78} & $1.23\times10^{-2}$ & $5.44\times10^{-2}$ \\
     \cline{2-4}\cline{7-8}
     \rule[-1mm]{0mm}{0.5cm}& sparse & $1-7\times10^{-16}$ & $5.31\times10^{-16}$ &  &  & $6.42\times10^{-3}$ & $4.66\times10^{-2}$ \\
     \cline{1-8}
     \rule[-1mm]{0mm}{0.5cm}\multirow{2}*{4}& dense & $1-2\times10^{-15}$ & $6.12\times10^{-16}$ & \multirow{2}*{105} & \multirow{2}*{160} & $1.42\times10^{-2}$ & $8.13\times10^{-2}$ \\
     \cline{2-4}\cline{7-8}
     \rule[-1mm]{0mm}{0.5cm}& sparse & $1-2\times10^{-15}$ & $6.25\times10^{-16}$ &  &  & $1.41\times10^{-2}$ & $8.51\times10^{-2}$ \\
     \hline
     \rule[-1mm]{0mm}{0.5cm}\multirow{2}*{5}& dense & $1-2\times10^{-15}$ & $1.70\times10^{-15}$ & \multirow{2}*{166} & \multirow{2}*{276} & $6.32\times10^{-2}$ & $1.54\times10^{-1}$ \\
     \cline{2-4}\cline{7-8}
     \rule[-1mm]{0mm}{0.5cm}& sparse & $1-2\times10^{-15}$ & $8.88\times10^{-16}$ &  &  & $5.08\times10^{-2}$ & $1.69\times10^{-1}$ \\
     \hline
    \end{tabular}}
\caption{Preparation of random quantum states with real negative coefficients. The results refer to a sample of 20 states.}
\label{tab:results_random_negative}
\end{table}

Finally, not noticing any substantial differences between the previously analyzed categories, the novel QSP algorithm has been tested from 6 to 10 qubits for random states, both dense and sparse, with only complex coefficients. The results are reported in Table~\ref{tab:results_random_complex2}.
\begin{table}[htbp]
\centering
\resizebox{0.9\textwidth}{!}{
    \begin{tabular}{|>{\centering\arraybackslash}m{0.8cm}|>{\centering\arraybackslash}m{1.5cm}|>{\centering\arraybackslash}m{3.1cm}|>{\centering\arraybackslash}m{3.1cm}|>{\centering\arraybackslash}m{1.2cm}|>{\centering\arraybackslash}m{1.2cm}|>{\centering\arraybackslash}m{1.7cm}|>{\centering\arraybackslash}m{1.7cm}|}
     \hline
     \multicolumn{8}{|c|}{\textbf{Simulation results for random states with complex amplitudes}}\\
     \hline
     \hline
     $n$ & Type of state & Avg Fidelity $\sigma^2\in[10^{-31},10^{-30}]$ & Avg Trace Distance $\sigma^2\in[10^{-32},10^{-31}]$ & Depth & CNOT gates & Avg Time Classical (s) & Avg Time Quantum (s) \\
     \hline
     \rule[-1mm]{0mm}{0.5cm}\multirow{2}*{6}& dense & $1-2\times10^{-15}$ & $8.22\times10^{-16}$ & \multirow{2}*{246} & \multirow{2}*{510} & $1.43\times10^{-1}$ & $4.23\times10^{-1}$ \\
     \cline{2-4}\cline{7-8}
     \rule[-1mm]{0mm}{0.5cm}& sparse & $1-3\times10^{-15}$ & $8.99\times10^{-16}$ &  &  & $1.62\times10^{-1}$ & $4.30\times10^{-1}$ \\
     \cline{1-8}
     \rule[-1mm]{0mm}{0.5cm}\multirow{2}*{7}& dense & $1-3\times10^{-15}$ & $1.11\times10^{15}$ & \multirow{2}*{427} & \multirow{2}*{930} & $3.40\times10^{-1}$ & 2.22 \\
     \cline{2-4}\cline{7-8}
     \rule[-1mm]{0mm}{0.5cm}& sparse & $1-3\times10^{-15}$ & $1.21\times10^{-15}$ &  &  & $3.88\times10^{-1}$ & 2.23 \\
     \cline{1-8}
     \rule[-1mm]{0mm}{0.5cm}\multirow{2}*{8}& dense & $1-3\times10^{-15}$ & $1.32\times10^{-15}$ & \multirow{2}*{647} & \multirow{2}*{1748} & $9.74\times10^{-1}$ & $4.23\times10^{2}$ \\
     \cline{2-4}\cline{7-8}
     \rule[-1mm]{0mm}{0.5cm}& sparse & $1-3\times10^{-15}$ & $1.11\times10^{-15}$ &  &  & $8.98\times10^{-1}$ & $4.53\times10^{2}$ \\
     \hline
     \rule[-1mm]{0mm}{0.5cm}\multirow{2}*{9}& dense & $1-3\times10^{-15}$ & $1.66\times10^{-15}$ & \multirow{2}*{1046} & \multirow{2}*{3354} & 3.70 & $2.63\times10^3$ \\
     \cline{2-4}\cline{7-8}
     \rule[-1mm]{0mm}{0.5cm}& sparse & $1-3\times10^{-15}$ & $1.30\times10^{-15}$ &  &  & 3.59 & $2.58\times10^3$ \\
     \hline
     \rule[-1mm]{0mm}{0.5cm}\multirow{2}*{10}& dense & $1-3\times10^{-15}$ & $1.55\times10^{-15}$ & \multirow{2}*{1871} & \multirow{2}*{6486} & 14.4 & $1.97\times10^4$ \\
     \cline{2-4}\cline{7-8}
     \rule[-1mm]{0mm}{0.5cm}& sparse & $1-3\times10^{-15}$ & $1.52\times10^{-15}$ &  &  & 14.9 & $2.06\times10^4$ \\
     \hline
    \end{tabular}}
\caption{Preparation of random quantum states with complex coefficients. The results refer to a sample of 20 states.}
\label{tab:results_random_complex2}
\end{table}

\begin{figure}[htbp]
\centering
\begin{tikzpicture}
\begin{axis}[
    width=0.9\textwidth,
    height=6cm,
    xlabel={Number of qubits, $n$},
    ylabel={Rotations - Phase Shifts},
    xmin=2, xmax=10,
    xtick={2,3,4,5,6,7,8,9,10},
    legend pos=south west,
    legend cell align=left,
    ymajorgrids]
\addplot+[mark=*] coordinates { (2,6) (3,14) (4,30) (5,62) (6,126) (7,254) (8,510) (9,1022) (10,2046) };
\addlegendentry{MOTT (original, PennyLane)}


\addplot+[mark=triangle*] coordinates { (2,-6) (3,-27) (4,-72) (5,-165) (6,-354) (7,-735) (8,-1500) (9,-3033) (10,-6102) };
\addlegendentry{SUN (original, 3 $\Lambda_n$)}

\addplot+[mark=diamond*] coordinates { (2,-5) (3,-16) (4,-39) (5,-86) (6,-181) (7,-372) (8,-755) (9,-1522) (10,-3057) };
\addlegendentry{OSUN (optimized, 1 $\Lambda_n$)}
\end{axis}
\end{tikzpicture}
\caption{Difference between rotations and phase shifts vs $n$}
\label{fig:rotations_vs_n}
\end{figure}
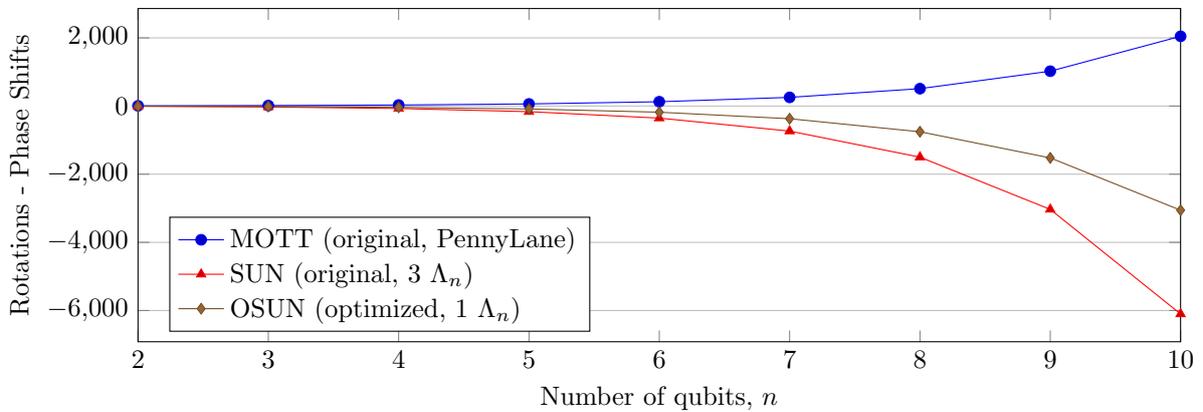

The QSP algorithm based on the new algebraic decomposition that uses a single $\Lambda$-type constructor for each UCG has been compared with the well-known algorithm by M\"ott\"onen et al.~\cite{mottonen2004transformation}, here denoted as MOTT, which is considered a reference standard for QSP without ancillary qubits. The comparison is meaningful because both algorithms are based on the traditional QSP structure defined by the UCG ladder sequence (Figure~\ref{fig:qsp_structure}). Some metrics are shown in Figures~\ref{fig:depth_vs_n}, \ref{fig:gates_vs_n}, \ref{fig:cnot_vs_n}, and \ref{fig:rotations_vs_n}. A comparison with the developments that followed the algorithm of M\"ott\"onen et al., such as those proposed by Bergholm~\cite{bergholm2005quantum}, by Plesch and Brukner~\cite{plesch2011quantum}, or with the generalizations introduced by Iten~\cite{iten2016quantum}, has not been covered in this work and will be the heart of a more systematic future benchmarking\footnote{The authors are aware of the recent result published by Carvalho et al.~\cite{de2025quantum} on the UCG class, which will also be compared}. The new decomposition proposed in this work, compared to MOTT, shows a significant reduction in depth already at the lower end of the first range of~\cite[Theorem 1]{sun2023asymptotically}, where the auxiliary register grows only polynomially ($m=2n$). Other points in Sun et al.'s parametric domain may show even greater reductions in light of this new decomposition. Therefore, comparison with M\"ott\"onen's current optimizations should be made once the optimal time-space trade-off is defined.
\begin{figure}[htbp]
\centering
\begin{tikzpicture}
\begin{axis}[
    width=0.9\textwidth,
    height=6cm,
    xlabel={Number of qubits, $n$},
    ylabel={Depth},
    xmin=2, xmax=10,
    xtick={2,3,4,5,6,7,8,9,10},
    legend pos=north west,
    legend cell align=left,
    ymajorgrids]
\addplot+[mark=*] coordinates { (2,8) (3,22) (4,52) (5,114) (6,240) (7,494) (8,1004) (9,2026) (10,4072) };
\addlegendentry{MOTT (original, PennyLane)}


\addplot+[mark=triangle*] coordinates { (2,60) (3,129) (4,222) (5,360) (6,549) (7,915) (8,1428) (9,2283) (10,3948) };
\addlegendentry{SUN (original, 3 $\Lambda_n$)}

\addplot+[mark=diamond*] coordinates { (2,39) (3,66) (4,105) (5,166) (6,246) (7,427) (8,647) (9,1046) (10,1871) };
\addlegendentry{OSUN (optimized, 1 $\Lambda_n$)}
\end{axis}
\end{tikzpicture}
\caption{Depth vs $n$}
\label{fig:depth_vs_n}
\end{figure}
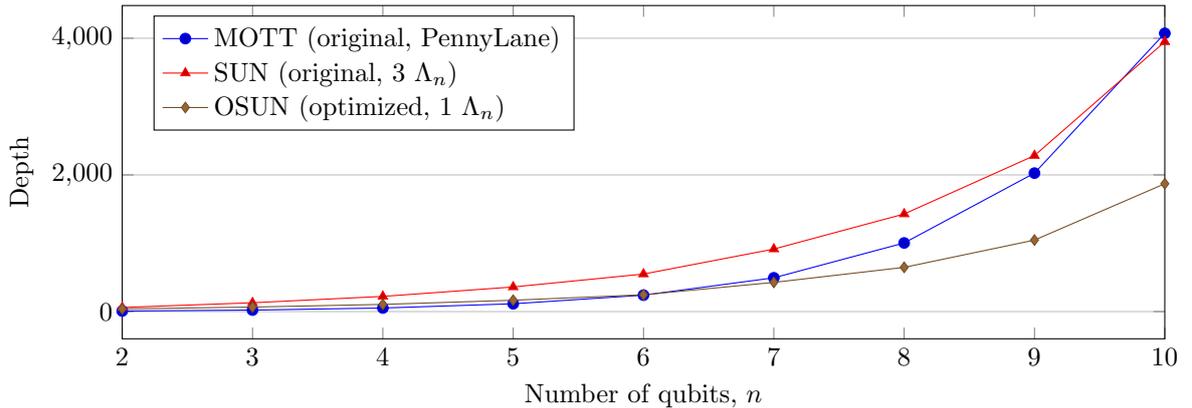
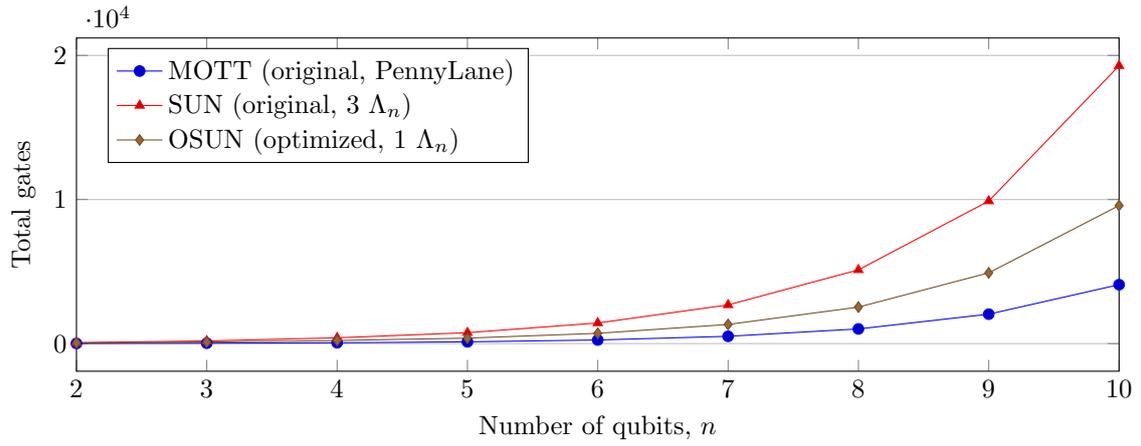
\begin{figure}[htbp]
\centering
\begin{tikzpicture}
\begin{axis}[
    width=0.9\textwidth,
    height=6cm,
    xlabel={Number of qubits, $n$},
    ylabel={Total gates},
    xmin=2, xmax=10,
    xtick={2,3,4,5,6,7,8,9,10},
    legend pos=north west,
    legend cell align=left,
    ymajorgrids]
\addplot+[mark=*] coordinates { (2,10) (3,26) (4,58) (5,122) (6,250) (7,506) (8,1018) (9,2042) (10,4090) };
\addlegendentry{MOTT (original, PennyLane)}


\addplot+[mark=triangle*] coordinates { (2,70) (3,185) (4,402) (5,757) (6,1430) (7,2685) (8,5116) (9,9893) (10,19284) };
\addlegendentry{SUN (original, 3 $\Lambda_n$)}

\addplot+[mark=diamond*] coordinates { (2,47) (3,104) (4,213) (5,380) (6,713) (7,1328) (8,2533) (9,4910) (10,9581) };
\addlegendentry{OSUN (optimized, 1 $\Lambda_n$)}
\end{axis}
\end{tikzpicture}
\caption{Total gates vs $n$}
\label{fig:gates_vs_n}
\end{figure}
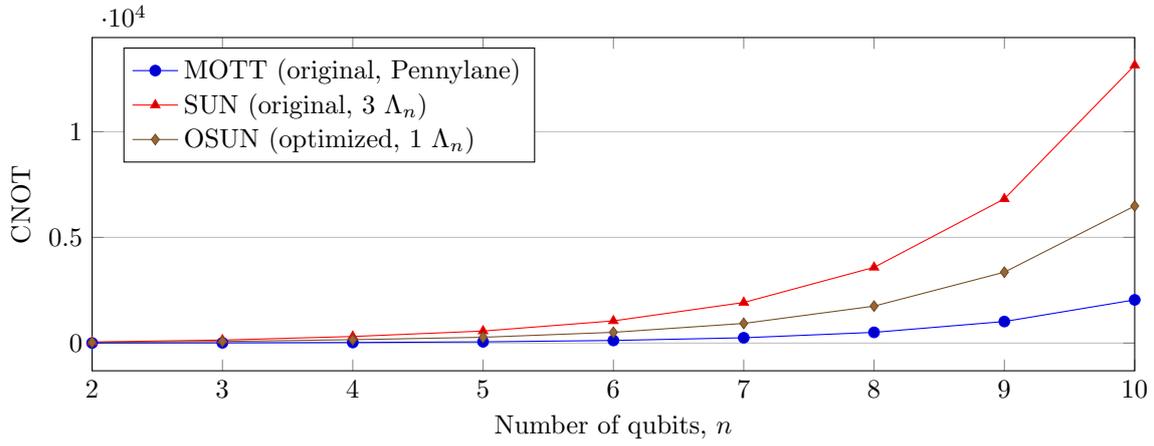
\begin{figure}[htbp]
\centering
\begin{tikzpicture}
\begin{axis}[
    width=0.9\textwidth,
    height=6cm,
    xlabel={Number of qubits, $n$},
    ylabel={CNOT},
    xmin=2, xmax=10,
    xtick={2,3,4,5,6,7,8,9,10},
    legend pos=north west,
    legend cell align=left,
    ymajorgrids]
\addplot+[mark=*] coordinates { (2,4) (3,12) (4,28) (5,60) (6,124) (7,252) (8,508) (9,1020) (10,2044) };
\addlegendentry{MOTT (original, Pennylane)}


\addplot+[mark=triangle*] coordinates { (2,54) (3,144) (4,312) (5,570) (6,1050) (7,1920) (8,3582) (9,6822) (10,13140) };
\addlegendentry{SUN (original, 3 $\Lambda_n$)}

\addplot+[mark=diamond*] coordinates { (2,36) (3,78) (4,160) (5,276) (6,510) (7,930) (8,1748) (9,3354) (10,6486) };
\addlegendentry{OSUN (optimized, 1 $\Lambda_n$)}
\end{axis}
\end{tikzpicture}
\caption{CNOT vs $n$}
\label{fig:cnot_vs_n}
\end{figure}

\section{Conclusions}\label{sec:conclusions}

In the traditional QSP framework based on successive UCG levels, the hypothesis of preparing the real part of the desired state separately from the complex part leads to an algebraic simplification. In fact, if the UCG ladder structure is entrusted with the task of preparing only the real moduli of each component of the target state, each UCG can be built with only $R_y$-blocks, which, in turn, can be implemented in terms of $R_z$, $H$, and $S$ gates. This reduction has a precise consequence for the QSP algorithm SUN, known in the literature for defining the optimal asymptotic space-time trade-off: for $2n\leqslant m\leqslant\frac{2^n}{n}$ ancillary qubits, each UCG can be implemented with one lambda operator (instead of 3), leading to systematic depth improvements. The complex part of the desired state is instead prepared with a single, maximum-dimensional lambda operator at the end of the UCG scheme. This new QSP algorithm, denoted as OSUN, does not change the complexity of SUN but improves the prefactors of the dominant terms, specifically by a factor of 3 in the linear term and by a factor of 2 in the exponential term. It is important to note that this algebraic reduction applies to the entire first parametric range of~\cite[Theorem 1]{sun2023asymptotically}, so future extensions of the algorithm to points in the domain where the ancillary register grows exponentially would benefit from the same optimization.

The OSUN algorithm has been implemented with the PennyLane library, and the code is available on GitHub~\cite{QSP-Sun}. The effectiveness of the algorithm has been verified in simulations involving up to 10 qubits for various types of target states. For each case, metrics such as depth, number of CNOTs, classical precomputation time, and quantum circuit simulation time have been reported. Furthermore, the new algorithm OSUN has been compared with the original version SUN and with the well-known M\"ott\"onen algorithm MOTT, which is an important reference as it is based on the same UCG scheme but does not use ancillary qubits. The results clearly show that OSUN has a shallower depth starting from 6 qubits, with a smoother growth curve compared to the original version, thanks to the optimization of the prefactors of the dominant terms. Regarding the total number of gates (and the number of CNOTs), OSUN does not improve upon MOTT. Note, however, that the gate set of CNOTs and rotations of MOTT is converted to a gate set of CNOTs and phase shifts in OSUN. Future developments could concern the extension of OSUN to other points in the parametric domain to better exploit the depth reductions afforded by the ancillae parallelization and systematic comparisons with the most recent implementations of the UCG class. Other topics to consider are qubit connectivity and how it affects performance on real hardware, or to what extent a translation in the Clifford+T gate set could affect the complexity.

\section*{Acknowledgments}
Giacomo Belli and Michele Amoretti acknowledge financial support from the European Union - NextGenerationEU, PNRR MUR project PE0000023-NQSTI. This research benefits from the High Performance Computing facility of the University of Parma, Italy (HPC.unipr.it).

\bibliographystyle{unsrt}
\bibliography{biblio}

@article{bergholm2018pennylane,
  title={Pennylane: Automatic differentiation of hybrid quantum-classical computations},
  author={Bergholm, Ville and Izaac, Josh and Schuld, Maria and Gogolin, Christian and Ahmed, Shahnawaz and Ajith, Vishnu and Alam, M Sohaib and Alonso-Linaje, Guillermo and AkashNarayanan, B and Asadi, Ali and others},
  journal={arXiv preprint arXiv:1811.04968},
  year={2018}
}

@book{nielsen2000quantum,
  title={Quantum Computation and Quantum Information},
  author={Nielsen, Michael A and Chuang, Isaac L},
  year={2000}
}

@article{grover2002creating,
  title={Creating superpositions that correspond to efficiently integrable probability distributions},
  author={Grover, Lov and Rudolph, Terry},
  journal={arXiv preprint quant-ph/0208112},
  year={2002}
}

@article{mottonen2004quantum,
  title={Quantum circuits for general multiqubit gates},
  author={M{\"o}tt{\"o}nen, Mikko and Vartiainen, Juha J and Bergholm, Ville and Salomaa, Martti M},
  journal={Physical review letters},
  volume={93},
  number={13},
  pages={130502},
  year={2004},
  publisher={APS}
}

@article{mottonen2004transformation,
  title={Transformation of quantum states using uniformly controlled rotations},
  author={Mottonen, Mikko and Vartiainen, Juha J and Bergholm, Ville and Salomaa, Martti M},
  journal={arXiv preprint quant-ph/0407010},
  year={2004}
}

@article{mottonen12006decompositions,
  title={Decompositions of general quantum gates},
  author={M{\"o}tt{\"o}nen, Mikko and Vartiainen, Juha J},
  journal={Trends in quantum computing research},
  pages={149},
  year={2006},
  publisher={Nova Publishers}
}

@article{bergholm2005quantum,
  title={Quantum circuits with uniformly controlled one-qubit gates},
  author={Bergholm, Ville and Vartiainen, Juha J and M{\"o}tt{\"o}nen, Mikko and Salomaa, Martti M},
  journal={Physical Review A—Atomic, Molecular, and Optical Physics},
  volume={71},
  number={5},
  pages={052330},
  year={2005},
  publisher={APS}
}

@article{plesch2011quantum,
  title={Quantum-state preparation with universal gate decompositions},
  author={Plesch, Martin and Brukner, {\v{C}}aslav},
  journal={Physical Review A—Atomic, Molecular, and Optical Physics},
  volume={83},
  number={3},
  pages={032302},
  year={2011},
  publisher={APS}
}

@article{zhang2021low,
  title={Low-depth quantum state preparation},
  author={Zhang, Xiao-Ming and Yung, Man-Hong and Yuan, Xiao},
  journal={Physical Review Research},
  volume={3},
  number={4},
  pages={043200},
  year={2021},
  publisher={APS}
}

@article{rosenthal2021query,
  title={Query and depth upper bounds for quantum unitaries via grover search},
  author={Rosenthal, Gregory},
  journal={arXiv preprint arXiv:2111.07992},
  year={2021}
}

@article{zhang2022quantum,
  title={Quantum state preparation with optimal circuit depth: Implementations and applications},
  author={Zhang, Xiao-Ming and Li, Tongyang and Yuan, Xiao},
  journal={Physical Review Letters},
  volume={129},
  number={23},
  pages={230504},
  year={2022},
  publisher={APS}
}

@article{sun2023asymptotically,
  title={Asymptotically optimal circuit depth for quantum state preparation and general unitary synthesis},
  author={Sun, Xiaoming and Tian, Guojing and Yang, Shuai and Yuan, Pei and Zhang, Shengyu},
  journal={IEEE Transactions on Computer-Aided Design of Integrated Circuits and Systems},
  volume={42},
  number={10},
  pages={3301--3314},
  year={2023},
  publisher={IEEE}
}

@article{yuan2023optimal,
  title={Optimal (controlled) quantum state preparation and improved unitary synthesis by quantum circuits with any number of ancillary qubits},
  author={Yuan, Pei and Zhang, Shengyu},
  journal={Quantum},
  volume={7},
  pages={956},
  year={2023},
  publisher={Verein zur F{\"o}rderung des Open Access Publizierens in den Quantenwissenschaften}
}

@article{yuan2023does,
  title={Does qubit connectivity impact quantum circuit complexity?},
  author={Yuan, Pei and Allcock, Jonathan and Zhang, Shengyu},
  journal={IEEE Transactions on Computer-Aided Design of Integrated Circuits and Systems},
  year={2023},
  publisher={IEEE}
}

@article{harrow2009quantum,
  title={Quantum algorithm for linear systems of equations},
  author={Harrow, Aram W and Hassidim, Avinatan and Lloyd, Seth},
  journal={Physical review letters},
  volume={103},
  number={15},
  pages={150502},
  year={2009},
  publisher={APS}
}

@article{iten2016quantum,
  title={Quantum circuits for isometries},
  author={Iten, Raban and Colbeck, Roger and Kukuljan, Ivan and Home, Jonathan and Christandl, Matthias},
  journal={Physical Review A},
  volume={93},
  number={3},
  pages={032318},
  year={2016},
  publisher={APS}
}

@article{vartiainen2004efficient,
  title={Efficient decomposition of quantum gates},
  author={Vartiainen, Juha J and M{\"o}tt{\"o}nen, Mikko and Salomaa, Martti M},
  journal={Physical review letters},
  volume={92},
  number={17},
  pages={177902},
  year={2004},
  publisher={APS}
}

@article{Iaconis2024,
	author = {Iaconis, Jason and Johri, Sonika and Zhu, Elton Yechao},
	journal = {npj Quantum Information},
	number = {1},
	pages = {15},
	title = {Quantum state preparation of normal distributions using matrix product states},
	volume = {10},
	year = {2024}
}

@article{sood2023towards,
  title={Towards quantum state preparation with materials science: An analytical review},
  author={Sood, Vaishali and Chauhan, Rishi Pal},
  journal={International Journal of Quantum Chemistry},
  volume={123},
  number={18},
  pages={e27148},
  year={2023},
  publisher={Wiley Online Library}
}

@article{lee2023evaluating,
  title={Evaluating the evidence for exponential quantum advantage in ground-state quantum chemistry},
  author={Lee, Seunghoon and Lee, Joonho and Zhai, Huanchen and Tong, Yu and Dalzell, Alexander M and Kumar, Ashutosh and Helms, Phillip and Gray, Johnnie and Cui, Zhi-Hao and Liu, Wenyuan and others},
  journal={Nature communications},
  volume={14},
  number={1},
  pages={1952},
  year={2023},
  publisher={Nature Publishing Group UK London}
}

@article{cao2019quantum,
  title={Quantum chemistry in the age of quantum computing},
  author={Cao, Yudong and Romero, Jonathan and Olson, Jonathan P and Degroote, Matthias and Johnson, Peter D and Kieferov{\'a}, M{\'a}ria and Kivlichan, Ian D and Menke, Tim and Peropadre, Borja and Sawaya, Nicolas PD and others},
  journal={Chemical reviews},
  volume={119},
  number={19},
  pages={10856--10915},
  year={2019},
  publisher={ACS Publications}
}

@article{berry2018improved,
  title={Improved techniques for preparing eigenstates of fermionic Hamiltonians},
  author={Berry, Dominic W and Kieferov{\'a}, M{\'a}ria and Scherer, Artur and Sanders, Yuval R and Low, Guang Hao and Wiebe, Nathan and Gidney, Craig and Babbush, Ryan},
  journal={npj Quantum Information},
  volume={4},
  number={1},
  pages={22},
  year={2018},
  publisher={Nature Publishing Group UK London}
}

@article{ward2009preparation,
  title={Preparation of many-body states for quantum simulation},
  author={Ward, Nicholas J and Kassal, Ivan and Aspuru-Guzik, Al{\'a}n},
  journal={The Journal of chemical physics},
  volume={130},
  number={19},
  year={2009},
  publisher={AIP Publishing}
}

@misc{QSP-Sun,
author = {Belli, Giacomo and Bersellini, Andrea and Amoretti, Michele},
title = {{Quantum State Preparation with $\Lambda$-type Operators (qsp-sun)}},
year = {2024},
howpublished = {https://github.com/qslab-unipr/qsp-sun}
}

@inproceedings{belli2025implementation,
  title={Implementation of an Optimally Bounded Algorithm for Quantum State Preparation},
  author={Belli, Giacomo and Bersellini, Andrea and Amoretti, Michele},
  booktitle={International Conference on Reversible Computation},
  pages={37--53},
  year={2025},
  organization={Springer}
}

@inproceedings{belli2025srbb,
  title={SRBB-Based Quantum State Preparation},
  author={Belli, Giacomo and Mordacci, Marco and Amoretti, Michele},
  booktitle={Proceedings of the 22nd ACM International Conference on Computing Frontiers},
  pages={172--175},
  year={2025}
}

@article{morales2024quantum,
  title={Quantum linear system solvers: A survey of algorithms and applications},
  author={Morales, Mauro ES and Pira, Lirand{\"e} and Schleich, Philipp and Koor, Kelvin and Costa, Pedro and An, Dong and Aspuru-Guzik, Al{\'a}n and Lin, Lin and Rebentrost, Patrick and Berry, Dominic W},
  journal={arXiv preprint arXiv:2411.02522},
  year={2024}
}

@article{bravo2023variational,
  title={Variational quantum linear solver},
  author={Bravo-Prieto, Carlos and LaRose, Ryan and Cerezo, Marco and Subasi, Yigit and Cincio, Lukasz and Coles, Patrick J},
  journal={Quantum},
  volume={7},
  pages={1188},
  year={2023},
  publisher={Verein zur F{\"o}rderung des Open Access Publizierens in den Quantenwissenschaften}
}

@article{low2019hamiltonian,
  title={Hamiltonian simulation by qubitization},
  author={Low, Guang Hao and Chuang, Isaac L},
  journal={Quantum},
  volume={3},
  pages={163},
  year={2019},
  publisher={Verein zur F{\"o}rderung des Open Access Publizierens in den Quantenwissenschaften}
}

@article{babbush2018encoding,
  title={Encoding electronic spectra in quantum circuits with linear T complexity},
  author={Babbush, Ryan and Gidney, Craig and Berry, Dominic W and Wiebe, Nathan and McClean, Jarrod and Paler, Alexandru and Fowler, Austin and Neven, Hartmut},
  journal={Physical Review X},
  volume={8},
  number={4},
  pages={041015},
  year={2018},
  publisher={APS}
}

@article{mcardle2020quantum,
  title={Quantum computational chemistry},
  author={McArdle, Sam and Endo, Suguru and Aspuru-Guzik, Al{\'a}n and Benjamin, Simon C and Yuan, Xiao},
  journal={Reviews of Modern Physics},
  volume={92},
  number={1},
  pages={015003},
  year={2020},
  publisher={APS}
}

@article{fomichev2024initial,
  title={Initial state preparation for quantum chemistry on quantum computers},
  author={Fomichev, Stepan and Hejazi, Kasra and Zini, Modjtaba Shokrian and Kiser, Matthew and Fraxanet, Joana and Casares, Pablo Antonio Moreno and Delgado, Alain and Huh, Joonsuk and Voigt, Arne-Christian and Mueller, Jonathan E and others},
  journal={PRX Quantum},
  volume={5},
  number={4},
  pages={040339},
  year={2024},
  publisher={APS}
}

@article{wang2022state,
  title={State preparation boosters for early fault-tolerant quantum computation},
  author={Wang, Guoming and Sim, Sukin and Johnson, Peter D},
  journal={Quantum},
  volume={6},
  pages={829},
  year={2022},
  publisher={Verein zur F{\"o}rderung des Open Access Publizierens in den Quantenwissenschaften}
}

@article{berry2025rapid,
  title={Rapid Initial-State Preparation for the Quantum Simulation of Strongly Correlated Molecules},
  author={Berry, Dominic W and Tong, Yu and Khattar, Tanuj and White, Alec and Kim, Tae In and Low, Guang Hao and Boixo, Sergio and Ding, Zhiyan and Lin, Lin and Lee, Seunghoon and others},
  journal={PRX Quantum},
  volume={6},
  number={2},
  pages={020327},
  year={2025},
  publisher={APS}
}

@article{rath2024quantum,
  title={Quantum data encoding: A comparative analysis of classical-to-quantum mapping techniques and their impact on machine learning accuracy},
  author={Rath, Minati and Date, Hema},
  journal={EPJ Quantum Technology},
  volume={11},
  number={1},
  pages={72},
  year={2024},
  publisher={Springer Berlin Heidelberg}
}

@article{caro2021encoding,
  title={Encoding-dependent generalization bounds for parametrized quantum circuits},
  author={Caro, Matthias C and Gil-Fuster, Elies and Meyer, Johannes Jakob and Eisert, Jens and Sweke, Ryan},
  journal={Quantum},
  volume={5},
  pages={582},
  year={2021},
  publisher={Verein zur F{\"o}rderung des Open Access Publizierens in den Quantenwissenschaften}
}

@article{gui2024spacetime,
  title={Spacetime-efficient low-depth quantum state preparation with applications},
  author={Gui, Kaiwen and Dalzell, Alexander M and Achille, Alessandro and Suchara, Martin and Chong, Frederic T},
  journal={Quantum},
  volume={8},
  pages={1257},
  year={2024},
  publisher={Verein zur F{\"o}rderung des Open Access Publizierens in den Quantenwissenschaften}
}

@article{malz2024preparation,
  title={Preparation of matrix product states with log-depth quantum circuits},
  author={Malz, Daniel and Styliaris, Georgios and Wei, Zhi-Yuan and Cirac, J Ignacio},
  journal={Physical Review Letters},
  volume={132},
  number={4},
  pages={040404},
  year={2024},
  publisher={APS}
}

@article{mao2024toward,
  title={Toward optimal circuit size for sparse quantum state preparation},
  author={Mao, Rui and Tian, Guojing and Sun, Xiaoming},
  journal={Physical Review A},
  volume={110},
  number={3},
  pages={032439},
  year={2024},
  publisher={APS}
}

@misc{ibm_scaling,
  author       = {IBM},
  title        = {How IBM will build the world's first large-scale, fault-tolerant quantum computer},
  howpublished = {\url{https://www.ibm.com/quantum/blog/large-scale-ftqc}},
  year         = {2025}
}

@misc{ibm_roadmap,
  author       = {IBM},
  title        = {Quantum Roadmap},
  howpublished = {\url{https://www.ibm.com/roadmaps/quantum/}},
  year         = {2025}
}

@article{khattar2025rise,
  title={Rise of conditionally clean ancillae for efficient quantum circuit constructions},
  author={Khattar, Tanuj and Gidney, Craig},
  journal={Quantum},
  volume={9},
  pages={1752},
  year={2025},
  publisher={Verein zur F{\"o}rderung des Open Access Publizierens in den Quantenwissenschaften}
}

@inproceedings{chatterjee2025quantum,
  title={Quantum Prometheus: Defying Overhead with Recycled Ancillas in Quantum Error Correction},
  author={Chatterjee, Avimita and Ghosh, Archisman and Ghosh, Swaroop},
  booktitle={2025 26th International Symposium on Quality Electronic Design (ISQED)},
  pages={1--7},
  year={2025},
  organization={IEEE}
}

@article{bravyi2024high,
  title={High-threshold and low-overhead fault-tolerant quantum memory},
  author={Bravyi, Sergey and Cross, Andrew W and Gambetta, Jay M and Maslov, Dmitri and Rall, Patrick and Yoder, Theodore J},
  journal={Nature},
  volume={627},
  number={8005},
  pages={778--782},
  year={2024},
  publisher={Nature Publishing Group UK London}
}

@article{mottonen2005transformation,
  title={Transformation of quantum states using uniformly controlled rotations},
  author={Mottonen, M and Vartiainen, JJ and Bergholm, V and Salomaa, MM},
  journal={Quantum Information and Computation},
  volume={5},
  number={6},
  pages={467--473},
  year={2005},
  publisher={Rinton Press}
}

@article{shende2006synthesis,
  title={Synthesis of quantum-logic circuits},
  author={Shende, VV and Bullock, SS and Markov, IL},
  journal={IEEE Transactions on Computer-Aided Design of Integrated Circuits and Systems},
  volume={25},
  number={6},
  pages={1000--1010},
  year={2006},
  publisher={IEEE Press Piscataway, NJ, USA}
}

@article{sun2021asymptotically,
  title={Asymptotically Optimal Circuit Depth for Quantum State Preparation and General Unitary Synthesis},
  author={Sun, Xiaoming and Tian, Guojing and Yang, Shuai and Yuan, Pei and Zhang, Shengyu},
  journal={arXiv preprint arXiv:2108.06150},
  year={2021}
}

@article{de2025quantum,
  title={Quantum multiplexer simplification for state preparation},
  author={de Carvalho, Jos{\'e} Alex and Batista, Carlos and de Veras, Tiago and Araujo, Israel and da Silva, Adenilton Jos{\'e}},
  journal={ACM Transactions on Quantum Computing},
  volume={6},
  number={4},
  pages={1--12},
  year={2025},
  publisher={ACM New York, NY}
}

\end{document}